\documentclass[showpacs,aps,amssymb,floatfix,prd,amsmath,preprintnumbers]{revtex4}
\usepackage{epstopdf}
\usepackage{capt-of}
\usepackage{graphicx}  
\usepackage{dcolumn}   
\usepackage{bm}
\begin{document}
\input epsf.tex

\title{\bf Magnetized strange quark matter in $f(R,T)$ gravity with bilinear and special form of time varying  deceleration parameter}

\author{
P.K. Sahoo\footnote{Department of Mathematics, Birla Institute of
Technology and Science-Pilani, Hyderabad Campus, Hyderabad-500078,
India,  Email:  pksahoo@hyderabad.bits-pilani.ac.in}, Parbati
Sahoo\footnote{Department of Mathematics, Birla Institute of
Technology and Science-Pilani, Hyderabad Campus, Hyderabad-500078,
India,  Email:  sahooparbati1990@gmail.com}, Binaya K. Bishi\footnote{Department of Mathematics, Visvesvaraya National Institute of Technology, Nagpur-440010,India,  Email:  binaybc@gmail.com}, Sezgin Ayg\"{u}n\footnote{Department of Physics, Canakkale Onsekiz Mart University, Arts and Sciences Faculty, Terzioglu Campus, 17020, Turkey, Email: saygun@comu.edu.tr}}

\affiliation{ }

\begin{abstract}
In this paper, we have studied homogeneous and anisotropic locally rotationally symmetric (LRS) Bianchi type-I model with magnetized strange quark matter (MSQM) distribution and cosmological constant $\Lambda$ in $f(R,T)$ gravity where $R$ is the Ricci scalar and $T$ the trace of matter source. The exact solutions of the field equations are obtained under bilinear and special form of time varying deceleration parameter (DP). Firstly, we have considered two specific forms of bilinear DP with a single parameter of the form: $q=\frac{\alpha(1-t)}{1+t}$ and $q=-\frac{\alpha t}{1+t}$, which leads to the constant or linear nature of the function based on the constant $\alpha$. Second one is the special form of the DP as $q=-1+\frac{\beta}{1+a^{\beta}}$. From the results obtained here, one can observe that in the early universe magnetic flux has more effects and it reduces gradually in the later stage. For  $t\rightarrow \infty$, we get $p\rightarrow -B_c$ and $\rho\rightarrow B_c$. The behaviour of strange quark matter along with magnetic epoch gives an idea of accelerated expansion of the universe as per the observations of the type Ia Supernovae.
\end{abstract}



\maketitle

\section{Introduction}\label{sec-I}

The theoretical arguments for the late-time cosmic acceleration are being a major issue among cosmologist of the twentieth century. The idea of this accelerated expansion of universe was discovered nearly 20 years ago by observations through Supernovae Ia \cite{riess98, perlmutter99, garnvich98, Perlmutter97,letelier83}, CMB \cite{spergel03, spergel07}, baryon acoustic oscillation (BAO) in galaxy clustering \cite{Eisenstein05, Percival07, Kazin14} and WMAP \cite{permutter03} etc. To investigate the nature of the universe, modern cosmology continues to test the above predictions, which leads to the refinement of cosmological models. The nature and behaviour of some unknown mechanism of the universe are responsible for this accelerated expansion, commonly referred as \textit{dark energy} and contains more energy budget of the universe along with negative pressure. It triggers one of the important issue to study the current acceleration of the universe.\\

Instead of resorting the mysterious concept of dark energy, there is an alternative way to reproduce the dynamics of the expanding universe through modified theories of gravity which is an extension of general relativity. These modifications can occur in several ways such as: one can use a base as the torsional formulation of general relativity called the teleparallel gravity equivalent to general relativity \cite{Einstein28} e.g. $f(T)$ gravity, where $T$ is the torsion scalar. On the other hand one can start the curvature formulation of general relativity in Einstein-Hilbert action by replacing the Ricci scalar with its arbitrary functions or even more complicated curvature invariants such as: $f(R)$, $f(R,G)$ gravity, and the latest $f(R,T)$ gravity proposed by Harko et al. \cite{Harko2011}. The matter Lagrangian of $f(R,T)$ gravity coupled with Ricci scalar $R$ and trace of energy-momentum tensor $T$. Such Lagrangian with matter content will differ from the Einstein's one. It has much significance to study late-time cosmic acceleration as well as dark energy and dark matter problem \cite{Nojiri07, Sotirou10, Nojiri11, Capozziello11}. Hypothetically matter plays more fundamental role in the description of gravitational effects of the universe. In literature, several models have been observed through different physically reliable matter in $f(R,T)$ gravity \cite{Sahoo14, Zubair16, Sahoo16, Sahoo17}.
Using Noether symmetry Momeni et al. \cite{Momeni2015} have researched flat FRW universe model with $f(R)$ and $f(R,T)$ gravitation theories. Baffou et al. \cite{Baffou2017} have investigated viscous chaplygin gas matter distribution in FRW universe model in $f(R,T)$ gravity. The compact stars in $f(R,T)$ theory have studied by Das et al. \cite{Das2016} using conformal killing vectors and the EoS $p=w\rho$ for fluid distribution. Alhamzawi and Alhamzawi \cite{Alhamzawi2016} have investigated the effects of $f(R,T)$ theory on gravitational lensing and also they compared the solutions with general relativity. Moraes and his co-authors have researched the details of $f(R,T)$ gravitation theory using the second law of thermodynamics \cite{Moraes1}, stellar equilibrium of compact stars \cite{Moraes2} and non-trivial polynomial function solutions of modified field equations \cite{Moraes3}. The observations indicate that various spiral galaxies, the Milky way, and pulsars have magnetic fields. These are considerable and so significant instrument in defining the structure of the early universe  \cite{Yilmaz2011}. But, there is still no precise information about the source of the magnetic fields \cite{Yilmaz2011}. Agrawal and Pawar \cite{Agrawal2017} have studied Bianchi type-V universe model with magnetized domain walls in $f(R,T)$ theory. Rani et al. have obtained magnetized string model (MSM) solutions for Bianchi type III universe in $f(R,T)$ theory \cite{Sarita2015}. Ram and Chandel have found dynamics of MSM in $f(R,T)$ theory for Bianchi type-V metric \cite{Ram2015}. Ayg\"{u}n et al. \cite{Aygun2016} have studied strange quark matter distribution for Marder type anisotropic universe model in $f(R,T)$ theory with $\Lambda$. Akta\c{s} and Ayg\"{u}n investigated the dynamics of MSQM distribution in FRW universe with reconstructed $f(R,T)$ gravity \cite{Aktas17}. Withal, because of the homogeneous and anisotropic cosmological models define the large scale structure of the universe in its early stages, spatially homogeneous and anisotropic universe models are so significant. In general near the singular point Bianchi type-I universe behave like Kasner universe. It has been discussed that a universe filled with matter, the initial anisotropy in Bianchi type-I universe quickly dies away and evolves into a FRW universe. It has simple mathematical structure and attracts researchers because of the ability to explain the cosmic evolution of the early universe. Therefore, the investigation of homogeneous and anisotropic Bianchi universe models are considerable in general relativity and other gravitational theories \cite{Nath2016}. In this context, Sahoo and Sivakumar \cite{Sahoo/2015} have studied $f(R,T)$ theory in LRS Bianchi type-I universe and they have presented the big rip singularity in this theory. Mishra et al. \cite{Chand2016} have investigated homogeneous and anisotropic Bianchi type-II universe for dark energy with/without a magnetic field in $f(R,T)$ gravitation theory. \c{C}aglar and Ayg\"{u}n \cite{Caglar2017} have obtained homogeneous and anisotropic Bianchi type-I solutions in $f(R,T)$ gravity with quark matter and $\Lambda$. Adhav \cite{Adhav2012} studied LRS Bianchi type-I universe model in $f(R,T)$ theory.\\

In this study, we have considered the magnetized strange quark matter distribution for LRS Bianchi type-I model of the universe in the framework of $f(R,T)$ gravity with cosmological constant $(\Lambda)$. Strange quark matter is composed of de-confined $u,d,$ and $s$ quarks, and treated as the ground state of matter as well as a strongly interacting matter, well described in \cite{Bodmer71, Witten84, Farhi84}. Recently, the effects of the magnetic field on the stability and on interacting properties of strange quark matter have attracted much attention \cite{Miransky15}. In 2016, Chang-Feng Li et al. \cite{Chang16} proposed an interaction between a magnetic field with strange quark matter in the framework of Nambu-Jona-Lasinio model. According to the literature, a new extensive idea about quantum chromodynamics theory (QCD) under some extreme conditions can be specified through the behaviour of the interaction coupling constant and the structure of dense matter \cite{Buballa05}. So many phenomenological models are used to analyze magnetized strange quark matter in literature \cite{Wen12, Chu15, Isayev13, Felipe08, Huang10}. One of the most successful phenomenological models for quark confinement is "MIT" bag model \cite{Chodos74}. By using a density independent bag pressure in quark confinement, the equation of state for SQM is considered as:
\begin{equation}\label{1}
p_m=\frac{1}{3}(\rho_m-4B_c)
\end{equation}
where, $B_c$ is called bag constant and its unit is $Mev(fm)^{-3}$. It has been expressed widely in different range values. Chakraborty et al. \cite{Chakraborty14} have described the range of $B_c$ in between $60-80$ MeV(fm)$^{-3}$. Again at zero temp, the value of $B_c$ is (150 MeV)$^4$ in $\beta$ equilibrium \cite{Fraga14}. Here, we considered the value of $B_c$ as $60$ MeV(fm)$^{-3}$. This conventional MIT bag model is used in literature for strange quark matter in a magnetic field, and moreover it was confirmed that there is an anisotropy pressure due to the presence of magnetic field \cite{Chakrabarty96, Martnez05}. This article is organized as follows: Section \ref{sec-II} outlines the basic formalism of $f(R,T)$ gravity and the corresponding field equations. Section \ref{sec-III} analyses the solution of field equations in details. Results and the observational behaviours of the model and the conclusion are discussed in Section \ref{sec-IV}.

\section{Field equations in $f(R,T)$ gravity}\label{sec-II}

By considering the metric dependent Lagrangian density $L_{matter}$, the respective field equation for $f(R,T)$ gravity are formulated from the Hilbert-Einstein variational principle in the following manner:
\begin{equation}\label{2}
S=\int \sqrt{-g}\biggl(\frac{1}{16\pi G}f(R,T)+L_{matter}\biggr)d^{4}x
\end{equation}%
where, $L_{matter}$ is the usual matter Lagrangian density of matter source, $f(R,T)$ is an arbitrary function of
Ricci scalar $R$ and the trace $T$ of the energy-momentum tensor $T_{ij}$ of
the matter source, and $g$ is the determinant of the metric tensor $g_{ij}$. The energy-momentum tensor $T_{ij}$ from Lagrangian matter is defined in the form:
\begin{equation}\label{3}
T_{ij}=-\frac{2}{\sqrt{-g}}\frac{\delta (\sqrt{-g}L_{matter})}{\delta g^{ij}}
\end{equation}%
and its trace is $T=g^{ij}T_{ij}$. Here, we have assumed that the matter Lagrangian $L_{matter}$ depends only on the
metric tensor component $g_{ij}$ rather than its derivatives. Hence, we
obtain:
\begin{equation}\label{4}
T_{ij}=g_{ij}L_{matter}-\frac{\partial L_{matter}}{\partial g^{ij}}
\end{equation}%
The $f(R,T)$ gravity field equations in general:
\begin{multline}\label{5}
f_{R}(R,T)\biggl(R_{ij}-\frac{1}{3}Rg_{ij}\biggr)+\frac{1}{6}f(R,T)g_{ij}= \\
8\pi -f_{T}(R,T)\biggl(T_{ij}-\frac{1}{3}Tg_{ij}\biggr)-f_{T}(R,T)\biggl(%
\Theta _{ij}-\frac{1}{3}\Theta g_{ij}\biggr)+\nabla _{i}\nabla _{j}f_{R}(R,T)
\end{multline}%
It is worth to mention here that the physical nature of the matter field through $%
\Theta _{ij}$ is used to form the field equations of $f(R,T)$ gravity.
There are  individual set of field equations for different frames of $f(R,T)$ gravity, here we consider one of them i.e $f(R,T)=R+2f(T)$ and the  field equation is given as:   \\
\begin{equation}\label{6}
R_{ij}-\frac{1}{2}Rg_{ij}=8\pi T_{ij}-2f'(T)T_{ij}-2f'(T)\Theta_{ij}+f(T)g_{ij}
\end{equation}
We consider the spatially homogeneous LRS Bianchi type-I metric as
\begin{equation}\label{7}
ds^{2}=dt^{2}-A^{2}dx^{2}-B^2(dy^{2}+dz^{2})
\end{equation}
where $A, B$ are functions of cosmic time $t$ only.\\
The energy momentum tensor for magnetized strange quark matters is considered as \cite{CGT1997, JDB2007}
\begin{equation}\label{8}
T_{ij}=(\rho+p+h^2)u_{i}u_{j}+\bigg(\frac{h^2}{2}-p \bigg)g_{ij}-h_i h_j
\end{equation}
where $u^i=(0,0,0,1)$ is the four velocity vector in co-moving coordinate system satisfying $u_iu_j=1$
and the magnetic flux $h^2$ is chosen in the $x$-direction satisfying $h_iu^i=0$. Quantizing the flux along the $x$-direction we can have
the magnetic field in the $yz$ plane. Here, $p$ is the proper pressure and $\rho$ is the energy density.

The field equation (\ref{5}) with cosmological constant $\Lambda$ and $f(T)=\mu T$ can be written as
\begin{equation}\label{9}
G_{ij}=[8\pi+2\mu]T_{ij}+[\mu \rho-\mu p+2\mu h^2+\Lambda]g_{ij}
\end{equation}
where $\mu$ is an arbitrary constant. To understand the dynamic history of the universe some physical parameters has more significant behaviour with respect to cosmic time $t$ such as; Hubble Parameter $(H)$, scale factor $(a)$, and the dimensionless DP $(q)$. The set of field equations for the metric  (\ref{7}) with Hubble parameter are obtained as
\begin{eqnarray}
2H_1H_2+H_2^2=-(12\pi+5\mu)h^2-(8\pi+3\mu)\rho+\mu p-\Lambda\\\label{10}
2\dot{H_2}+3H_2^2= (4\pi-\mu)h^2+(8\pi+3\mu)p-\mu \rho-\Lambda\\ \label{11}
\dot{H_1}+\dot{H_2}+H_1^2+H_2^2+H_1H_2=-(4\pi+3\mu)h^2+(8\pi+3\mu)p-\mu \rho-\Lambda \label{12}
\end{eqnarray}
here $H_1=\frac{\dot{A}}{A}$ and $H_2=\frac{\dot{B}}{B}$ are the directional Hubble parameters with $H=\frac{H_1+2H_2}{3}$ is the mean Hubble parameter. The dot represent derivatives with respect to time $t$. For the metric (\ref{7}), the scalar expansion $\theta$ and shear scalar $\sigma$ are defined as follows
\begin{equation}\label{13}
\theta=3H=H_1+2H_2
\end{equation}
\begin{equation}\label{14}
\sigma^2=\frac{1}{3}(H_1-H_2)^2
\end{equation}

\section{Solutions of Field equations}\label{sec-III}

The set of field equations  (\ref{10}-\ref{12}) have six unknowns ($A, B, \rho, p, h^2$ and $\Lambda$ ) with three equations. In order to get physically viable models of the universe which are consistent with the observations, we have considered the following physically plausible relation.\\
\begin{enumerate}
\item Initially, we have assumed the linear relationship between the directional Hubble parameters $H_1$ and $H_2$ as
\begin{equation}\label{15}
H_1=nH_2
\end{equation}
where $n\geq 0$ is an arbitrary constant which takes care about the anisotropy nature of the model. The above equation yields the shear scalar $\sigma$ is proportional to the scalar expansion $\theta.$
\item Secondly, we have considered the equation of state (EoS) for strange quark matter as
\begin{equation}\label{16}
p=\frac{\rho-4B_c}{3}
\end{equation}
where $B_c$ is bag constant \cite{Sotani2004}.
\item Finally, we have assumed different time varying deceleration parameter $q$. Because the study of a various cosmological model with time-dependent deceleration parameter gives an extensive new look into modern cosmology in the account of the current accelerated expansion of the universe. Several cosmological models are constructed through constant deceleration parameter in earlier, which referred as Berman's law of constant DP \cite{Berman88}.
\end{enumerate}
Using the first assumption the field equation (10)-(12) take the form
\begin{eqnarray}
\frac{9(2n+1)}{(n+2)^2}H^2=-(12\pi+5\mu)h^2-(8\pi+3\mu)\rho+\mu p-\Lambda\\ \label{17}
\biggl[\frac{27}{(n+2)^2}-\frac{6(1+q)}{n+2}\biggr]H^2= (4\pi-\mu)h^2+(8\pi+3\mu)p-\mu \rho-\Lambda\\ \label{18}
\biggl[\frac{9(n^2+n+1)}{(n+2)^2}-\frac{3(n+1)(1+q)}{n+2}\biggr]H^2=-(4\pi+3\mu)h^2+(8\pi+3\mu)p-\mu \rho-\Lambda \label{19}
\end{eqnarray}
After using the EoS from equation (\ref{16}), we obtain the following values
\begin{equation}\label{20}
h^2=\frac{3(n-1)(q-2)}{2(4\pi+\mu)(n+2)}H^2
\end{equation}
\begin{equation}\label{21}
\rho=\frac{-3}{4(4\pi+\mu)}\biggl[\frac{9(n-1)}{(n+2)^2}+\frac{3(3+qn-2n)}{(n+2)}\biggr]H^2+B_c
\end{equation}
\begin{equation}\label{22}
p=\frac{-1}{4(4\pi+\mu)}\biggl[\frac{9(n-1)}{(n+2)^2}+\frac{3(3+qn-2n)}{(n+2)}\biggr]H^2-B_c
\end{equation}
\begin{equation}\label{23}
\Lambda=\biggl[\frac{3[(12n\pi+3n\mu-n^2 \mu+24\pi+10\mu)q]}{2(4\pi+\mu)(n+2)^2}+\frac{(-26\mu+18n\mu+6n^2 \mu-76\pi)}{2(4\pi+\mu)(n+2)^2}\biggr]H^2-(8\pi+4\mu)B_c
\end{equation}

The dimensionless deceleration parameter has great importance as it is responsible for the understanding of the evolution of the universe. The deceleration parameter is defined in terms of the scale factor and scale factor is a function of time. So it is always motives the researchers to investigate the time dependent deceleration parameter rather than the constant deceleration parameter. Thus we have interested to investigate the time dependent deceleration parameter.  For the investigation, two type of DP are considered i.e. bilinear DP and special form of DP. The considered form of bilinear deceleration parameters in model-I and model-II evolves into the super-exponential expansion phase unless $\alpha>1$. This observation may be considered as one of the possible fates of the universe with reference to the cosmological observations. Thus it is important to investigate the bilinear form of deceleration parameters. Along with the above the the special form of DP, shows phase transition for $\beta>1$ and acceleration for $\beta$ in the interval $(0,1]$.  Finally these three DP leads to accelerating models of our universe. Thus it makes sense to investigate the discussed DPs. Our previous work \cite{Sahoo017} deals with the linearly varying DP and the result is compatible with recent observations. Here, we have considered a bilinear DP \cite{Mishra16} in two forms i.e (i) $ q=\frac{\alpha(1-t)}{1+t}$ and (ii) $ q=\frac{-\alpha t}{1+t}$ and a time varying DP \cite{Debnath2009} i.e $q=-1+\frac{\beta}{1+a^{\beta}}$.\\

\subsection{Model-I: $q(t)=\frac{\alpha(1-t)}{1+t}$}

Here, we have considered the first form of the bilinear deceleration parameter \cite{Mishra16}
\begin{equation}\label{24}
q=-\frac{a\ddot{a}}{\dot{a}^2}=\frac{\alpha(1-t)}{1+t}
\end{equation}

where $a$ is the average scale factor and $\alpha>0$ is a constant. Again $q>0$ for $0<t<1$ and $q\leq0$ for $t\geq1$.\\
The Hubble's parameter $H=\frac{\dot{a}}{a}$ can be obtained from equation (\ref{24}) as
\begin{equation}\label{25}
H=\frac{1}{(1-\alpha)t+2\alpha \log(1+t)}
\end{equation}
Integrating (\ref{25}) we have
\begin{equation}\label{26}
a= a_0 t^{\frac{1}{1+\alpha}}e^{G(t)}
\end{equation}
where
\begin{equation*}
G(t)=\frac{\alpha}{(1+\alpha)^2}t+\frac{-2\alpha+\alpha^2}{6(1+\alpha)^3}t^2+\frac{3\alpha-2\alpha^2+\alpha^3}{18(1+\alpha)^4}t^3
+\frac{-18\alpha+11\alpha^2-14\alpha^3+2\alpha^4}{180(1+\alpha)^5}t^4+O(t^5)
\end{equation*}
and $a_0$ is the constant of integration. The values of $\rho, h^2 \ \&\ \Lambda$  are obtained as
\begin{equation}\label{27}
h^2=\frac{3(n-1)[(\alpha-2)-(\alpha+2)t]}{2(1+t)(4\pi+\mu)(n+2)}H^2
\end{equation}
\begin{equation}\label{28}
\rho= \frac{-3}{4(4\pi+\mu)}\biggl[\frac{9(n-1)}{(n+2)^2}+\frac{3[3+(\alpha-2)n+(3-3n)t]}{(1+t)(n+2)}\biggr]H^2+B_c
\end{equation}
\begin{equation}\label{29}
p= \frac{-1}{4(4\pi+\mu)}\biggl[\frac{9(n-1)}{(n+2)^2}+\frac{3[3+(\alpha-2)n+(3-3n)t]}{(1+t)(n+2)}\biggr]H^2-B_c
\end{equation}
\begin{equation}\label{30}
\Lambda=\biggl[\frac{3[(12n\pi+3n\mu-n^2 \mu+24\pi+10\mu)\alpha(1-t)]}{2(1+t)(4\pi+\mu)(n+2)^2}+\frac{(-26\mu+18n\mu+6n^2 \mu-76\pi)}{2(4\pi+\mu)(n+2)^2}\biggr]H^2-(8\pi+4\mu)B_c
\end{equation}
The expressions of other physical parameters of this model like as, spatial volume, expansion scalar, shear scalar and anisotropy parameter are given as
\begin{eqnarray}
V=a^3=a_0^3 t^{\frac{3}{1+\alpha}}e^{3G(t)}\\ \label{31}
\theta=3H=\frac{3}{(1-\alpha)t+2\alpha \log(1+t)}\\ \label{32}
\sigma^2=\frac{1}{2}(H_1^2+2H_2^2-\frac{\theta^2}{3})=\frac{3(n-1)^2}{(n+2)^2}[(1-\alpha)t+2\alpha \log(1+t)]^{-2} \\ \label{33}
\Delta=\frac{2(n-1)^2}{(n+2)^2} \label{34}
\end{eqnarray}

\begin{figure}[ht!]
\minipage{0.32\textwidth}
  \includegraphics[width=60mm]{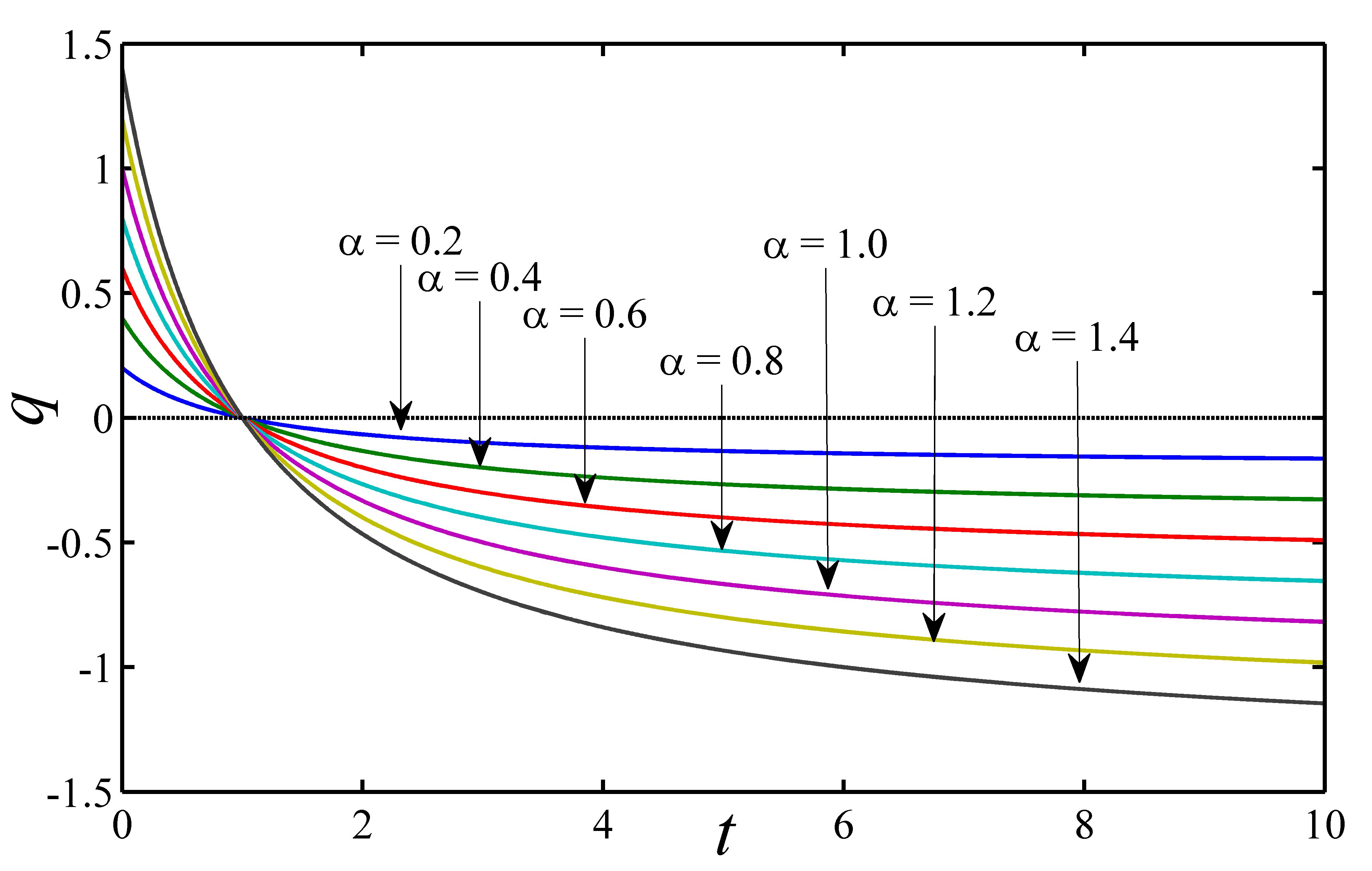}
  \caption{Variation of deceleration parameter against time for different $\alpha$}\label{fig1}
\endminipage\hfill
\minipage{0.32\textwidth}
  \includegraphics[width=60mm]{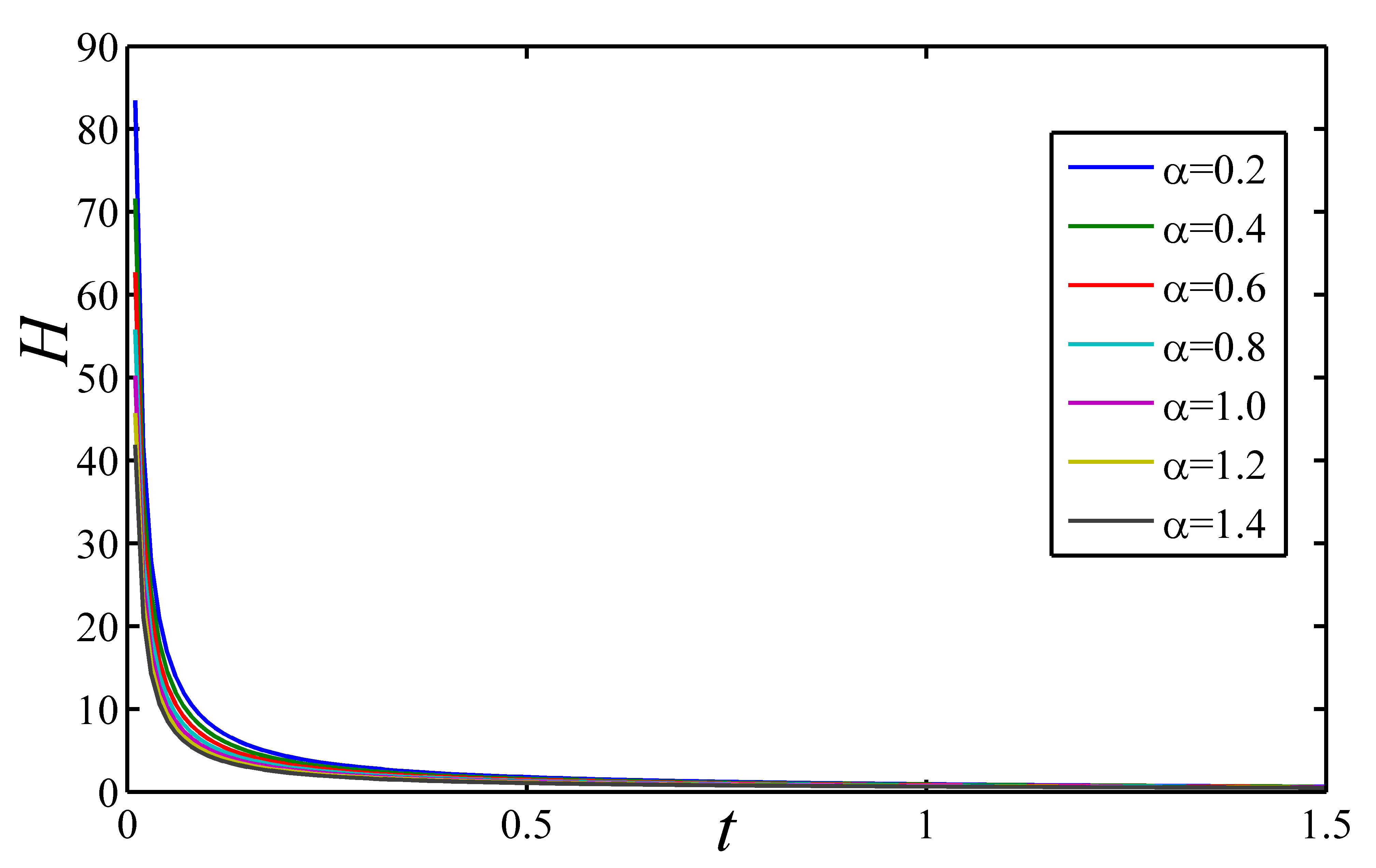}
  \caption{Variation of Hubble parameter against time for different $\alpha$}\label{fig2}
\endminipage\hfill
\minipage{0.32\textwidth}%
  \includegraphics[width=60mm]{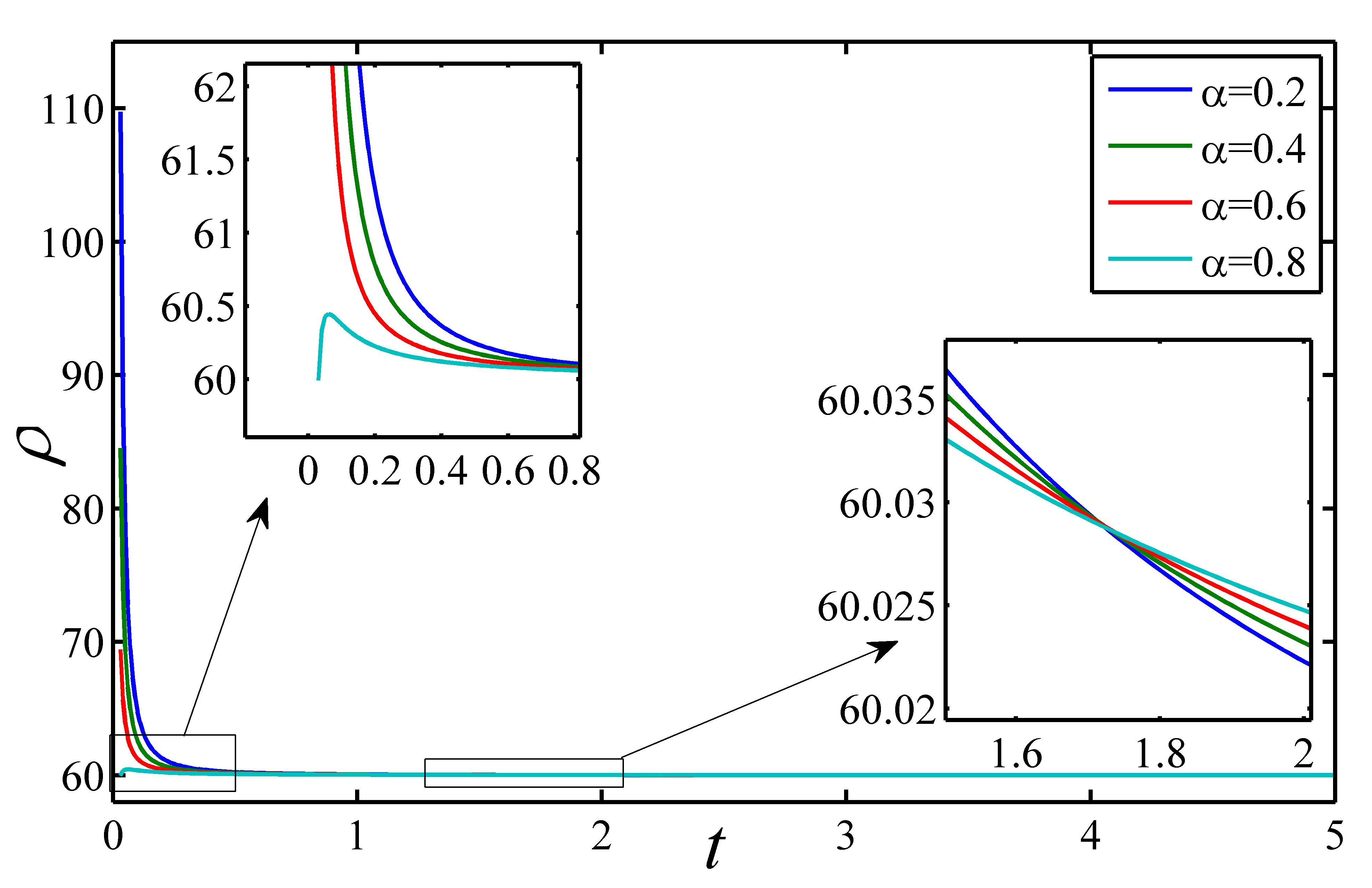}
  \caption{Variation of energy density against time for $\lambda=0.1$, $m=3.5$, $B_c=60$ and different $\alpha$ }\label{fig3}
\endminipage
\end{figure}

\begin{figure}[ht!]
\minipage{0.32\textwidth}
  \includegraphics[width=60mm]{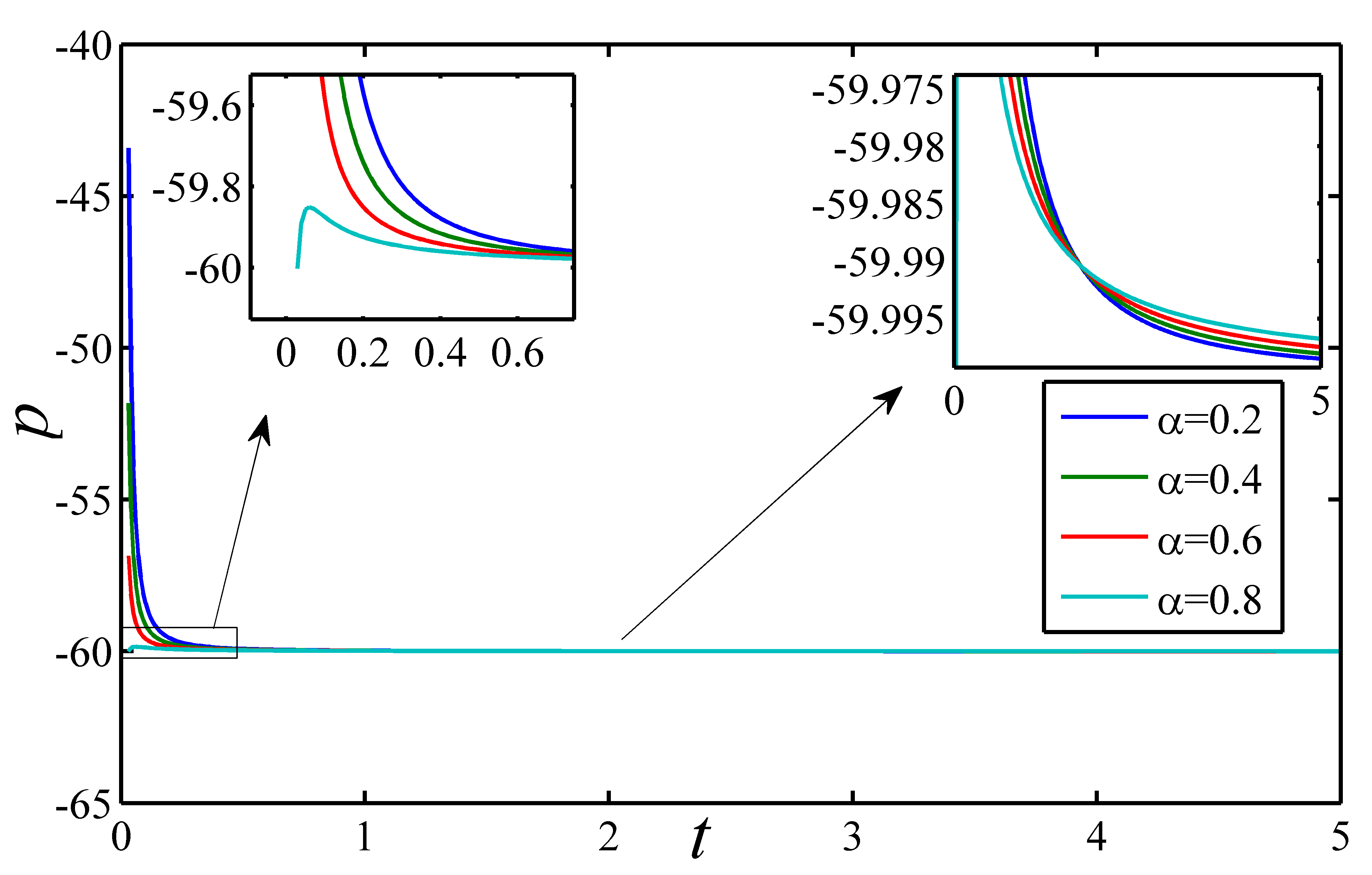}
  \caption{Variation of pressure parameter against time for $\lambda=0.1$, $m=3.5$, $B_c=60$ and different $\alpha$}\label{fig4}
\endminipage\hfill
\minipage{0.32\textwidth}
  \includegraphics[width=60mm]{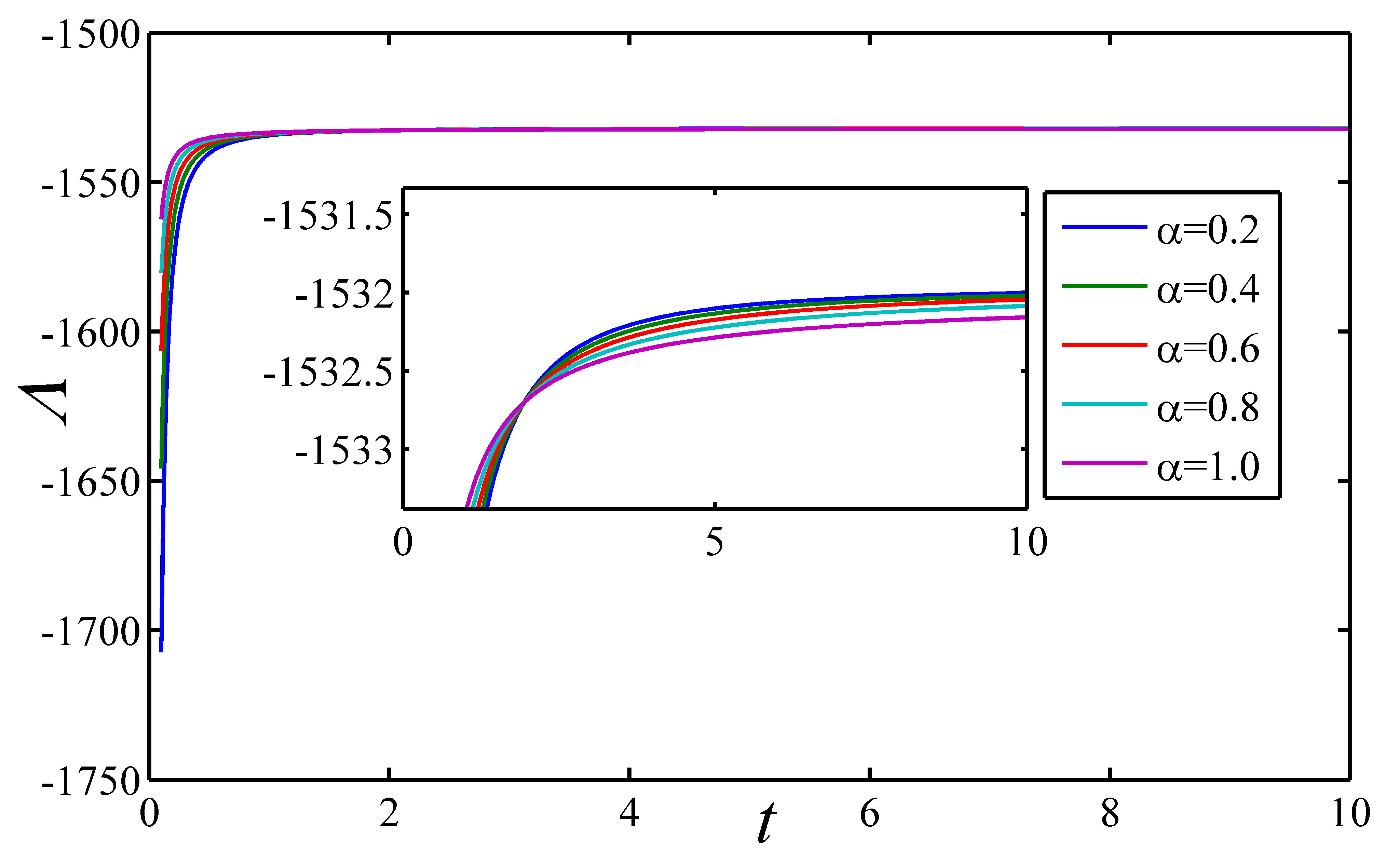}
  \caption{Variation of cosmological constant against time for $\lambda=0.1$, $m=0.2$, $B_c=60$ and different $\alpha$ }\label{fig5}
\endminipage\hfill
\minipage{0.32\textwidth}%
  \includegraphics[width=60mm]{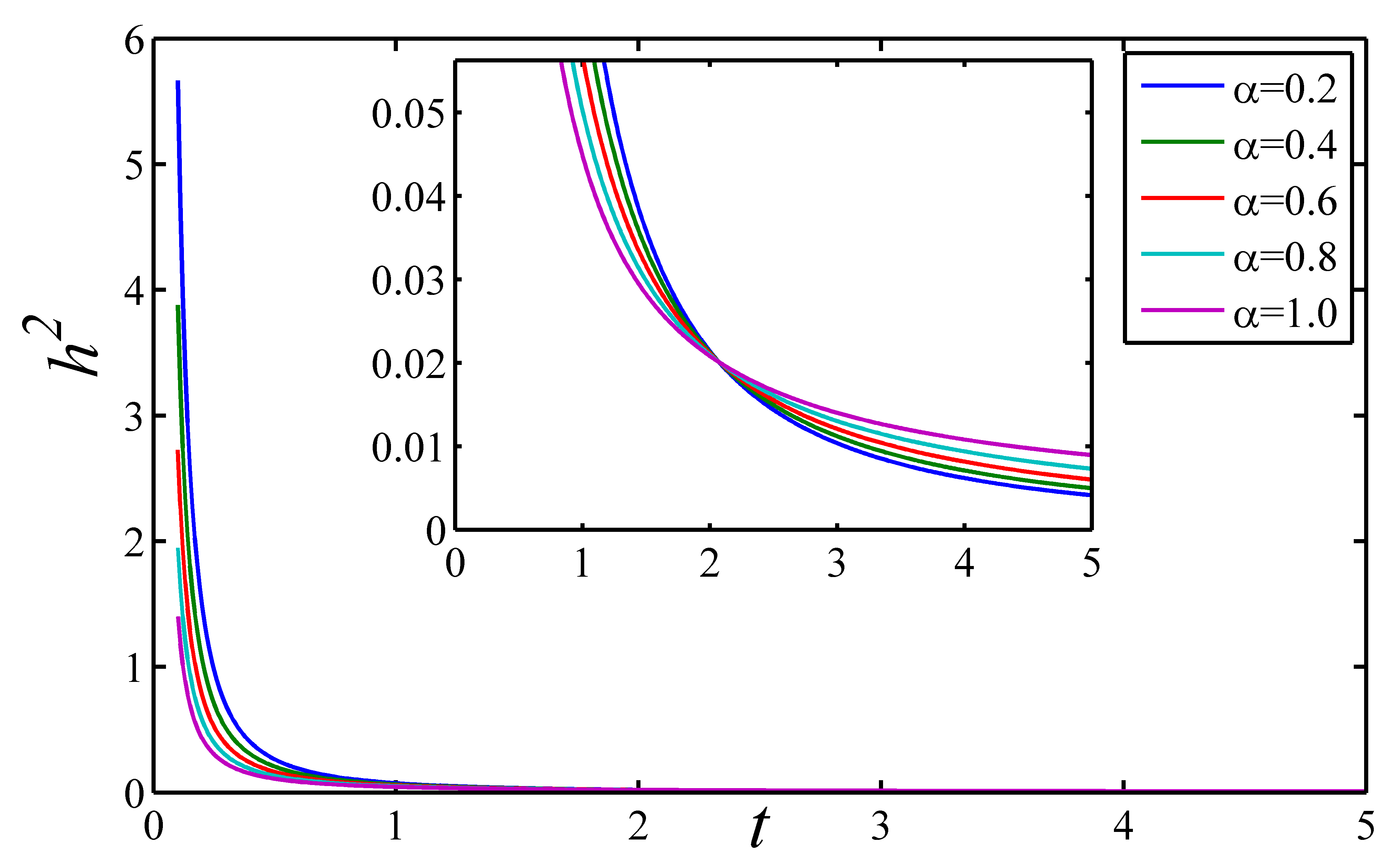}
  \caption{Variation of magnetic flux $h^2$ against time for $\lambda=0.1$, $m=0.2$ and different $\alpha$}\label{fig6}
\endminipage
\end{figure}

First set of figures (\ref{fig1}-\ref{fig6}) of model-I consists the quantitative behaviour of $q, H, \rho, p, \Lambda,$ and $h^2$. Fig \ref{fig1}  indicates that, in this model phase transition takes place  as $q$ is evolving with positive to negative valued for different $\alpha$ against time. Evolution of Hubble parameter against time with set of $\alpha$ values is presented in Fig \ref{fig2}. The Hubble parameter posses an initial singularity at $t=0$, then it tends to zero in later as $t$ tends to infinity.  Fig \ref{fig3} and Fig \ref{fig4} depicts the profile of energy density and pressure against time for different $\alpha$ respectively. From equations (\ref{28}) and (\ref{29}), one can notice that, $\rho\rightarrow B_c$ and $p\rightarrow -B_c$ when $t\rightarrow \infty$. We would like to point out that, the approach of $\rho$ toward $B_c$ is different for different $n$. For $n\in(0.9,3.5)$, $\rho\rightarrow B_c$ from the left of $B_c$ and for $n\geq 3.5$, $\rho\rightarrow B_c$  from the right of $B_c$.  As a representative case we choose $n=3.5$ and different $\alpha$ for energy density profile, which is presented in Fig \ref{fig3}. In case of pressure, for $n\in(0,16]$, pressure is purely negative valued function of time. In the interval $(0,2]$ and [3.5,16] of $n$, $p\rightarrow -B_c$  from the left and right of $B_c$ respectively. Similarly for energy density here also we choose $n=3.5$ and different $\alpha$ for representative case of pressure profile. Variation of cosmological constant against time is presented in Fig \ref{fig5}. From equation (\ref{30}), it is clear that $\lambda\rightarrow -(8\pi+4\mu)B_c$ when $t\rightarrow \infty$. Depending upon the value of $n$ and $\alpha$, the approach of $\Lambda$ toward $-(8\pi+4\mu)B_c$ is different means that $\lambda\rightarrow -(8\pi+4\mu)B_c$ from either side of the value of $-(8\pi+4\mu)B_c$. Here we noticed that, cosmological constant is negative quantity. As a representative case we choose $n=0.2$ and different $\alpha$ for profile of cosmological constant. The profile of magnetic flux $h^2$ is depicted in Fig \ref{fig6}. We have noticed that, $h^2$ is positive valued for $n\in (0,1)$ and negative valued for $n>1$ for provided values of $\alpha$. Further, at initial time $t=0$, the spatial volume $V$ is zero and gradually increases exponentially with time. It is interesting to note that, for $n\neq 1$, the model is anisotropic for late time and not free from shear whereas for $n=1$, it is isotropic and shear free.

\subsection{Model-II: $q(t)=-\frac{\alpha t}{1+t}$}

In this case we have considered the second form of the bilinear deceleration parameter \cite{Mishra16}
\begin{equation}\label{35}
q(t)=-\frac{\alpha t}{1+t}
\end{equation}
and it yields the Hubble parameter as
\begin{equation}\label{36}
H=\frac{1}{(1-\alpha)t+\alpha log(1+t)}
\end{equation}
Integrating the above equation, we have
\begin{equation}\label{37}
a= a_0 t e^{F(t)}
\end{equation}
where
\begin{equation*}
F(t)=\frac{\alpha}{2}t+\frac{-4\alpha+3\alpha^2}{24}t^2+\frac{6\alpha-8\alpha^2+3\alpha^3}{72}t^3
+\frac{-144\alpha+260\alpha^2-180\alpha^3+45\alpha^4}{2880}t^4+O(t^5)
\end{equation*}
The values of $\rho$, $h^2$ and  $\Lambda$  are obtained as
\begin{equation}\label{38}
h^2=\frac{3(n-1)[-(\alpha+2)t-2]}{2(1+t)(4\pi+\mu)(n+2)}H^2
\end{equation}
\begin{equation}\label{39}
\rho= \frac{-3}{4(4\pi+\mu)}\biggl[\frac{9(n-1)}{(n+2)^2}+\frac{3[3-2n+(3-2n-\alpha n)t]}{(1+t)(n+2)}\biggr]H^2+B_c
\end{equation}
\begin{equation}\label{40}
p=\frac{-1}{4(4\pi+\mu)}\biggl[\frac{9(n-1)}{(n+2)^2}+\frac{3[3-2n+(3-2n-\alpha n)t]}{(1+t)(n+2)}\biggr]H^2-B_c
\end{equation}
\begin{equation}\label{41}
\Lambda=\biggl[\frac{3[(12n\pi+3n\mu-n^2 \mu+24\pi+10\mu)(-\alpha t)]}{2(1+t)(4\pi+\mu)(n+2)^2}\\+\frac{(-26\mu+18n\mu+6n^2 \mu-76\pi)}{2(4\pi+\mu)(n+2)^2}\biggr]H^2-(8\pi+4\mu)B_c
\end{equation}
The other physical parameters of this model are given as:
\begin{eqnarray}
V=a^3=a_0^3 t^3 e^{3F(t)}\\\label{42}
\theta=3H=\frac{3}{(1-\alpha)t+\alpha log(1+t)}\\\label{43}
\sigma^2=\frac{1}{2}(H_1^2+2H_2^2-\frac{\theta^2}{3})=\frac{3(n-1)^2}{(n+2)^2}[(1-\alpha)t+\alpha log(1+t)]^{-2} \\\label{44}
\Delta=\frac{2(n-1)^2}{(n+2)^2}\label{45}
\end{eqnarray}

\begin{figure}[ht]
\minipage{0.32\textwidth}
  \includegraphics[width=60mm]{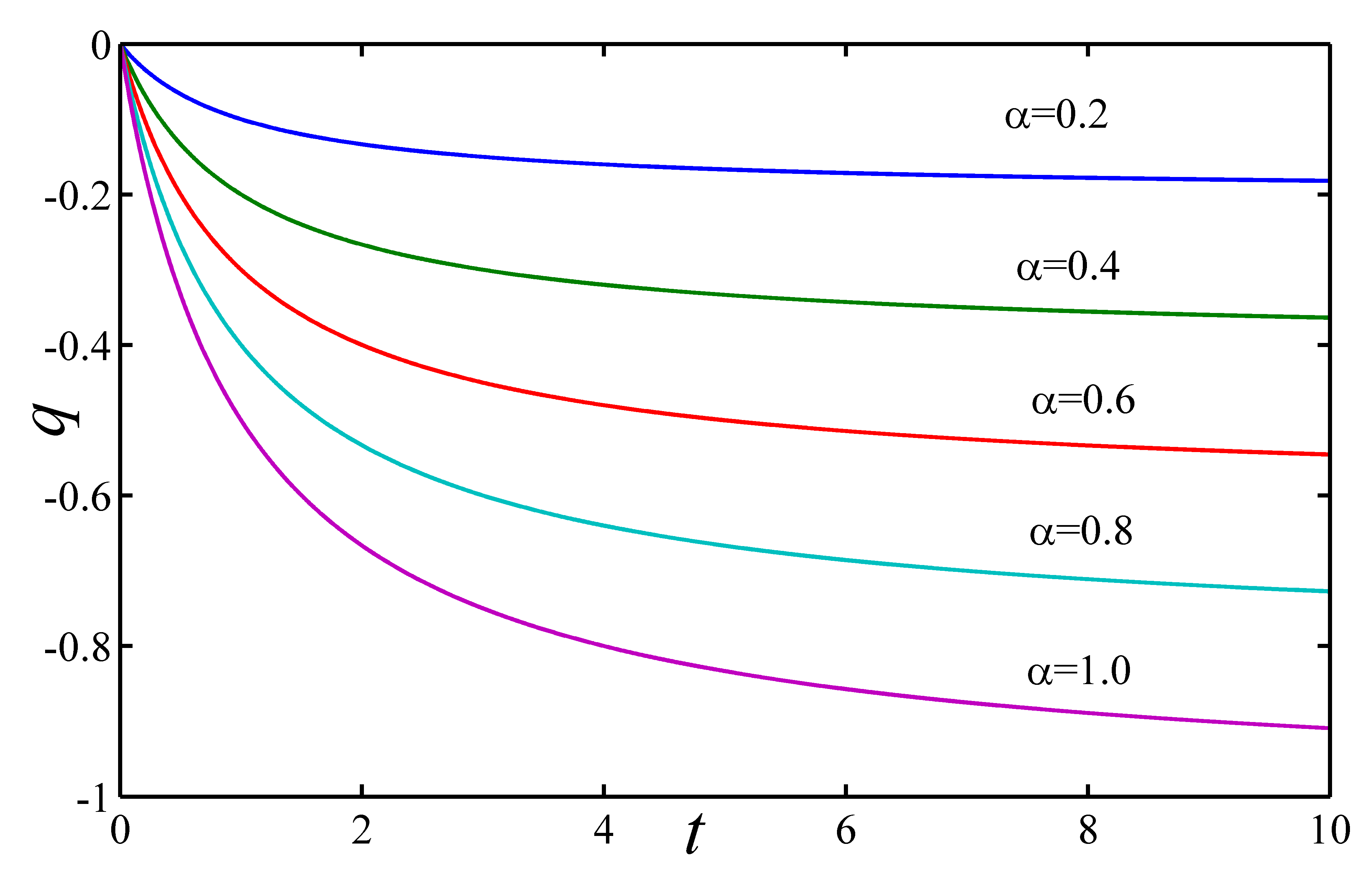}
  \caption{Variation of deceleration parameter against time for different $\alpha$}\label{fig7}
\endminipage\hfill
\minipage{0.32\textwidth}
  \includegraphics[width=60mm]{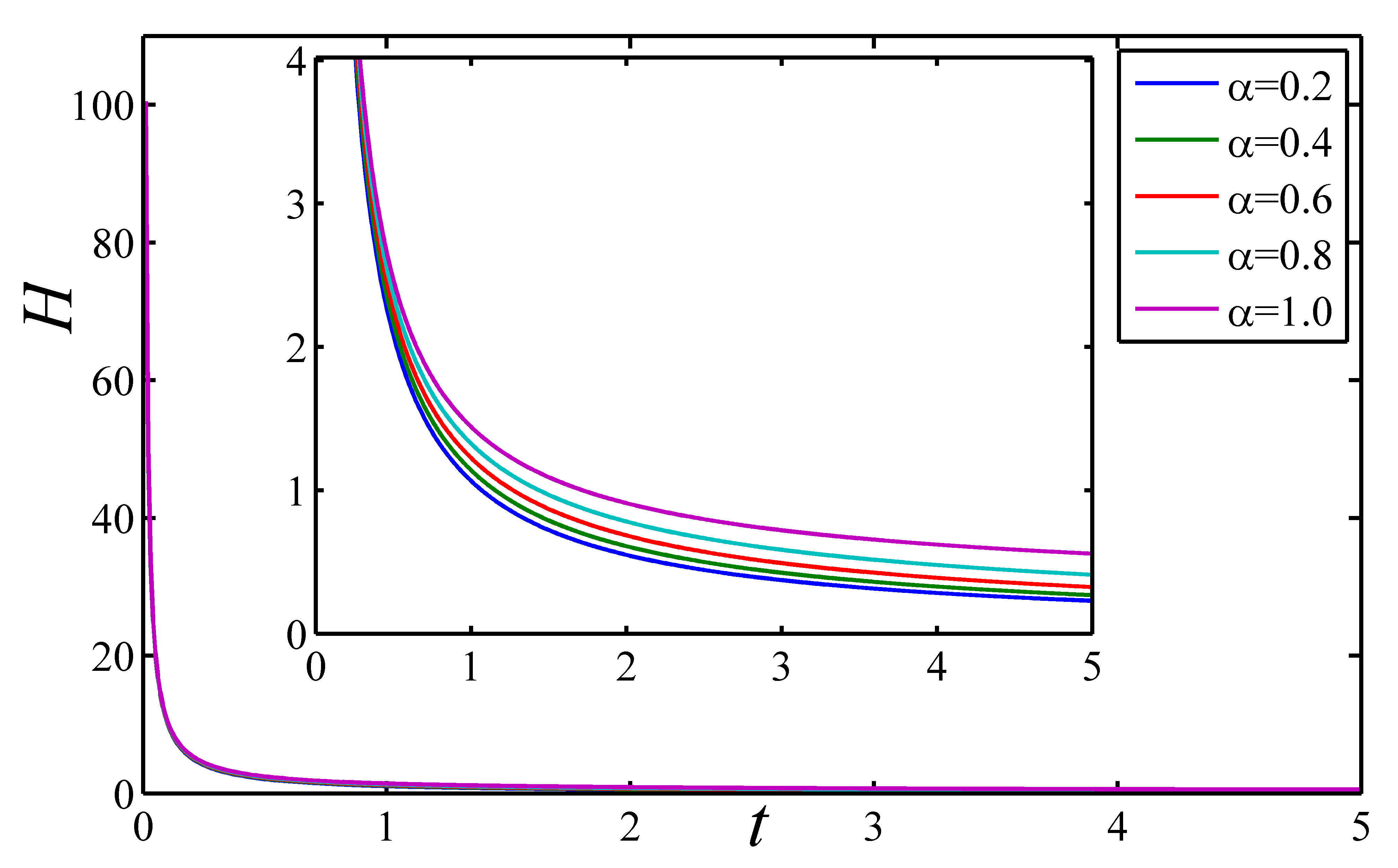}
  \caption{Variation of Hubble parameter against time for different $\alpha$}\label{fig8}
\endminipage\hfill
\minipage{0.32\textwidth}%
  \includegraphics[width=60mm]{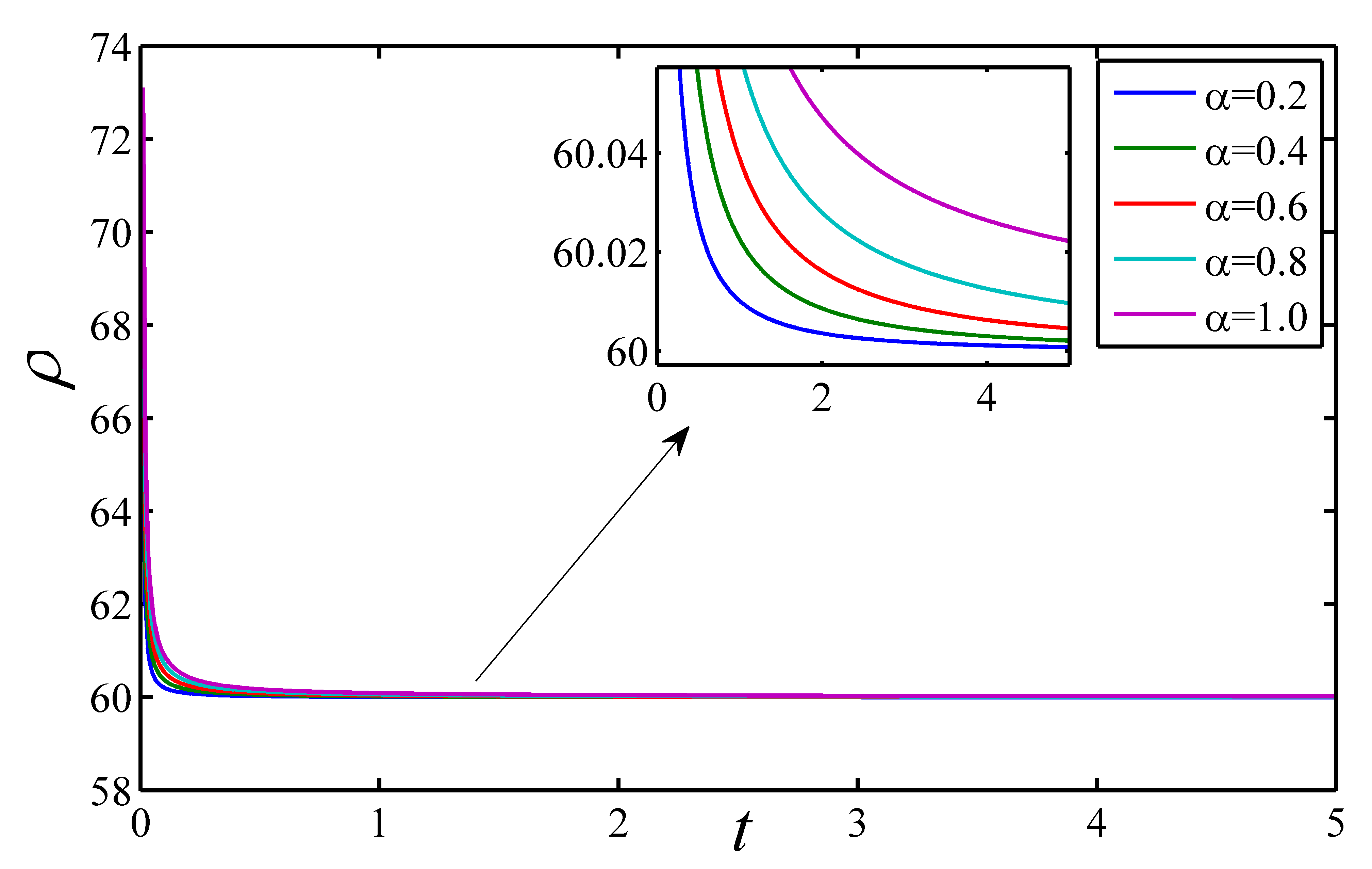}
  \caption{Variation of energy density against time for $\lambda=0.1$, $m=1.83$, $B_c=60$ and different $\alpha$}\label{fig9}
\endminipage
\end{figure}

\begin{figure}[ht]
\minipage{0.32\textwidth}
  \includegraphics[width=60mm]{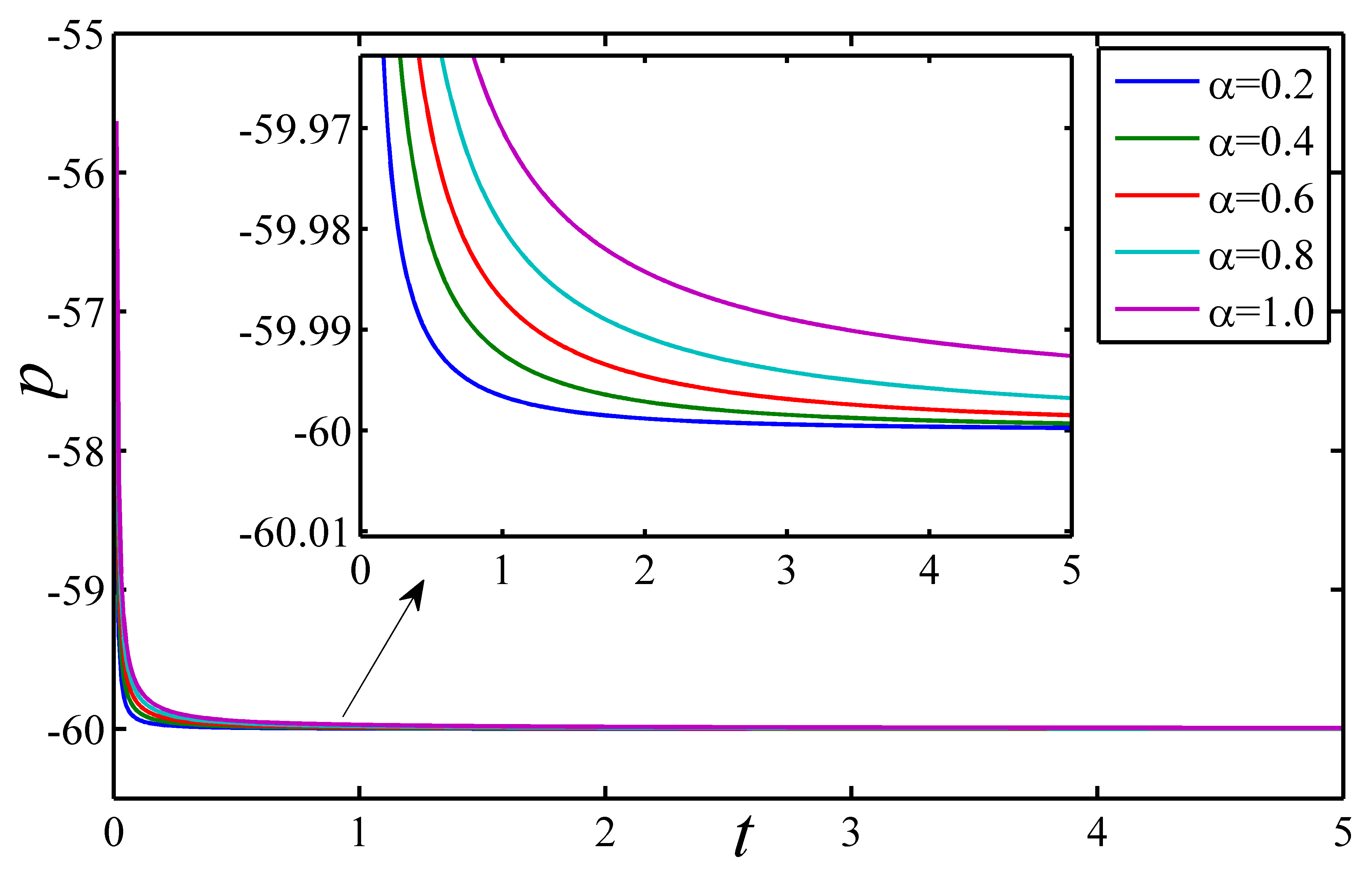}
  \caption{Variation of pressure parameter against time for $\lambda=0.1$, $m=1.83$, $B_c=60$ and different $\alpha$ }\label{fig10}
\endminipage\hfill
\minipage{0.32\textwidth}
  \includegraphics[width=60mm]{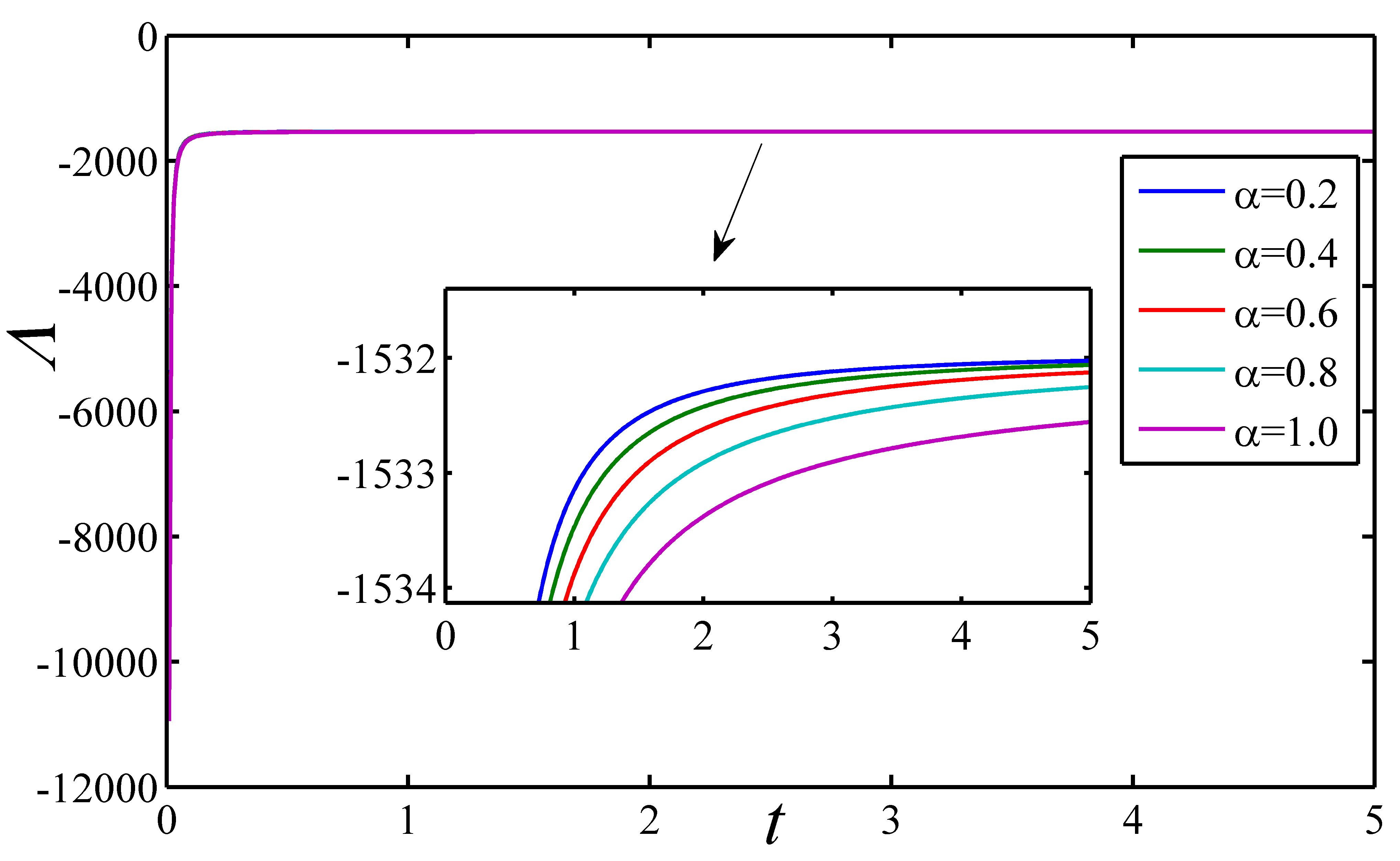}
  \caption{Variation of cosmological constant against time for $\lambda=0.1$, $m=1.83$, $B_c=60$ and different $\alpha$  }\label{fig11}
\endminipage\hfill
\minipage{0.32\textwidth}%
  \includegraphics[width=60mm]{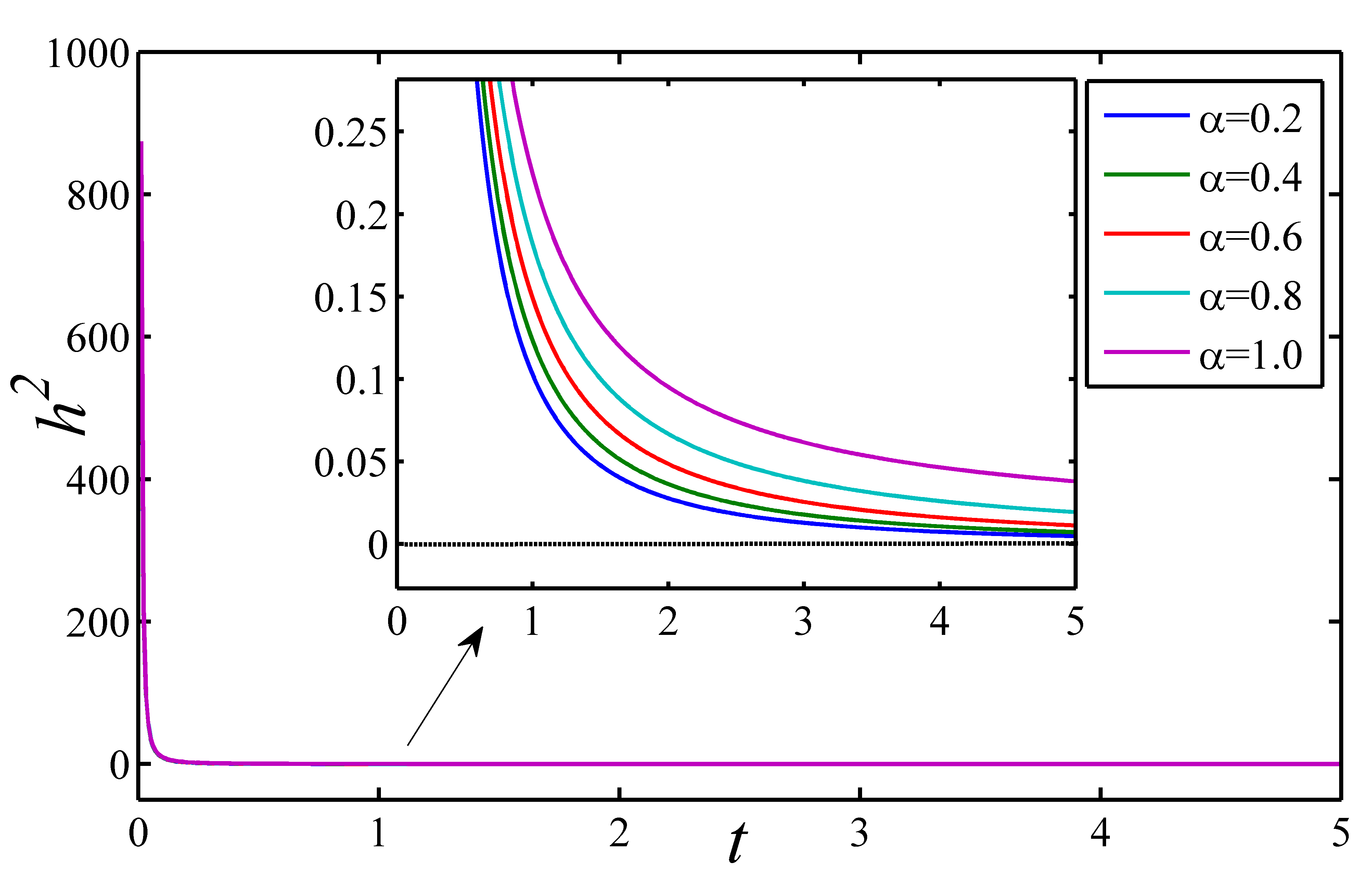}
  \caption{Variation of magnetic flux $h^2$ against time for $\lambda=0.1$, $m=0.2$, $B_c=60$ and different $\alpha$ }\label{fig12}
\endminipage
\end{figure}

Figure \ref{fig7} and Figure \ref{fig8} represent the variation of deceleration parameter and Hubble parameter against time for different $\alpha$. Here $q<0$ for $\alpha>0$. Thus in this model our Universe is accelerating. Moreover, specifically for $0<\alpha\leq 1$ \& $t>0$ $\Rightarrow$ $q\in (-1,0)$ and $\alpha>1$ \& $t>0$ $\Rightarrow$ $q\in (0,-2)$, our Universe is accelerating with exponential expansion (see Figure \ref{fig7}) and super exponential expansion respectively. Hubble parameter is decreasing with the time and tending to zero when time tends to infinity.
The positivity of energy density, $\forall \alpha \in (0,1)$ restrict the $n\geq 1.75$. The energy density $\rho\rightarrow B_c$ from the left of $B_c$ in the interval $n\in[1.75,1.82]$ and from the right of $B_c$ for $n>1.82$ (see figure \ref{fig10}). In case of profile of pressure, $p\rightarrow -B_c$ from the left of $B_c$ for $n\in[1.75,1.82]$ and $p\rightarrow -B_c$ from the right of $B_c$ for $n>1.82$ (see figure \ref{fig11}). Pressure is a negative valued function of time. The profile of cosmological constant depicted in the Figure \ref{fig10}. Here we noticed that, for provided value of $\alpha$ and $\forall n>0$, the cosmological constant is negative and $\Lambda\rightarrow -(8\pi+4\mu)B_c$ (See figure \ref{fig10} and equation \ref{40}). The other physical quantities like volume, shear scalar, expansion scalar, magnetic flux have the similar qualitative behaviour as that of model-I.

\subsection{Model-III: $q(t)=-1+\frac{\beta}{1+a^{\beta}}$}

In this case, we have considered the time special form of time varying deceleration parameter \cite{Debnath2009}
\begin{equation}\label{46}
q(t)=-1+\frac{\beta}{1+a^{\beta}}
\end{equation}
where $\beta >0$ is a constant. Consequently, the Hubble's parameter is
\begin{equation}\label{47}
H=A_1 (1+a^{-\beta})
\end{equation}
where $A_1$ is an integrating constant. Again integrating the above equation, we have
\begin{equation}\label{48}
a=(e^{A_1 \beta t}-1)^{\frac{1}{\beta}}
\end{equation}
The values of $\rho$, $h^2$ and $\Lambda$  are obtained as
\begin{equation}\label{49}
h^2=\frac{3(n-1)[-3+\beta e^{-A_1 \beta t}]}{2(4\pi+\mu)(n+2)}H^2
\end{equation}
\begin{equation}\label{50}
\rho= \frac{-3}{4(4\pi+\mu)}\biggl[\frac{9(n-1)}{(n+2)^2}+\frac{3[3+n\beta e^{-A_1 \beta t}-3n]}{(n+2)}\biggr]H^2+B_c
\end{equation}
\begin{equation}\label{51}
p= \frac{-1}{4(4\pi+\mu)}\biggl[\frac{9(n-1)}{(n+2)^2}+\frac{3[3+n\beta e^{-A_1 \beta t}-3n]}{(n+2)}\biggr]H^2-B_c
\end{equation}
\begin{equation}\label{52}
\Lambda=\biggl[\frac{3[(12n\pi+3n\mu-n^2 \mu+24\pi+10\mu)(-1+\beta e^{-A_1 \beta t})]}{2(4\pi+\mu)(n+2)^2}\\+\frac{(-26\mu+18n\mu+6n^2 \mu-76\pi)}{2(4\pi+\mu)(n+2)^2}\biggr]H^2-(8\pi+4\mu)B_c
\end{equation}
The remaining physical parameters are as follows:
\begin{eqnarray}
V=a^3= (e^{A_1 \beta t}-1)^\frac{1}{\beta}\\ \label{53}
\theta=3H=3A_1 e^{A_1 \beta t}(e^{A_1 \beta t}-1)^{-1}\\ \label{54}
\sigma^2=\frac{1}{2}(H_1^2+2H_2^2-\frac{\theta^2}{3})=\frac{3(n-1)^2}{(n+2)^2}[A_1^2 e^{2A_1 \beta t}(e^{A_1 \beta t}-1)^{-2}] \\ \label{55}
\Delta= 6\frac{\sigma^2}{\theta^2}=\frac{2(n-1)^2}{(n+2)^2}\label{56}
\end{eqnarray}

\begin{figure}[ht]
\minipage{0.32\textwidth}
  \includegraphics[width=60mm]{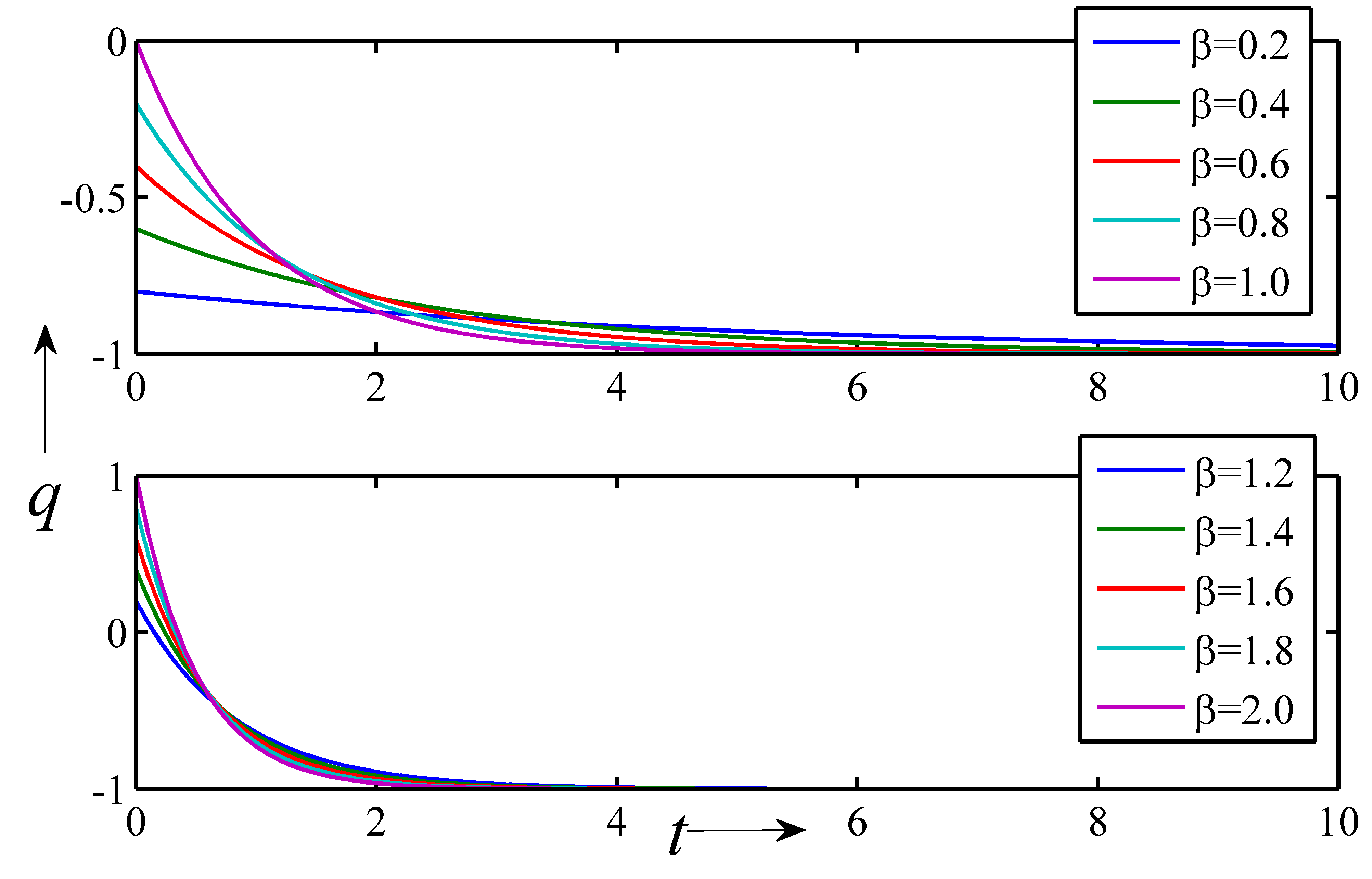}
  \caption{Variation of deceleration parameter against time for different $\beta$ }\label{fig13}
\endminipage\hfill
\minipage{0.32\textwidth}
  \includegraphics[width=60mm]{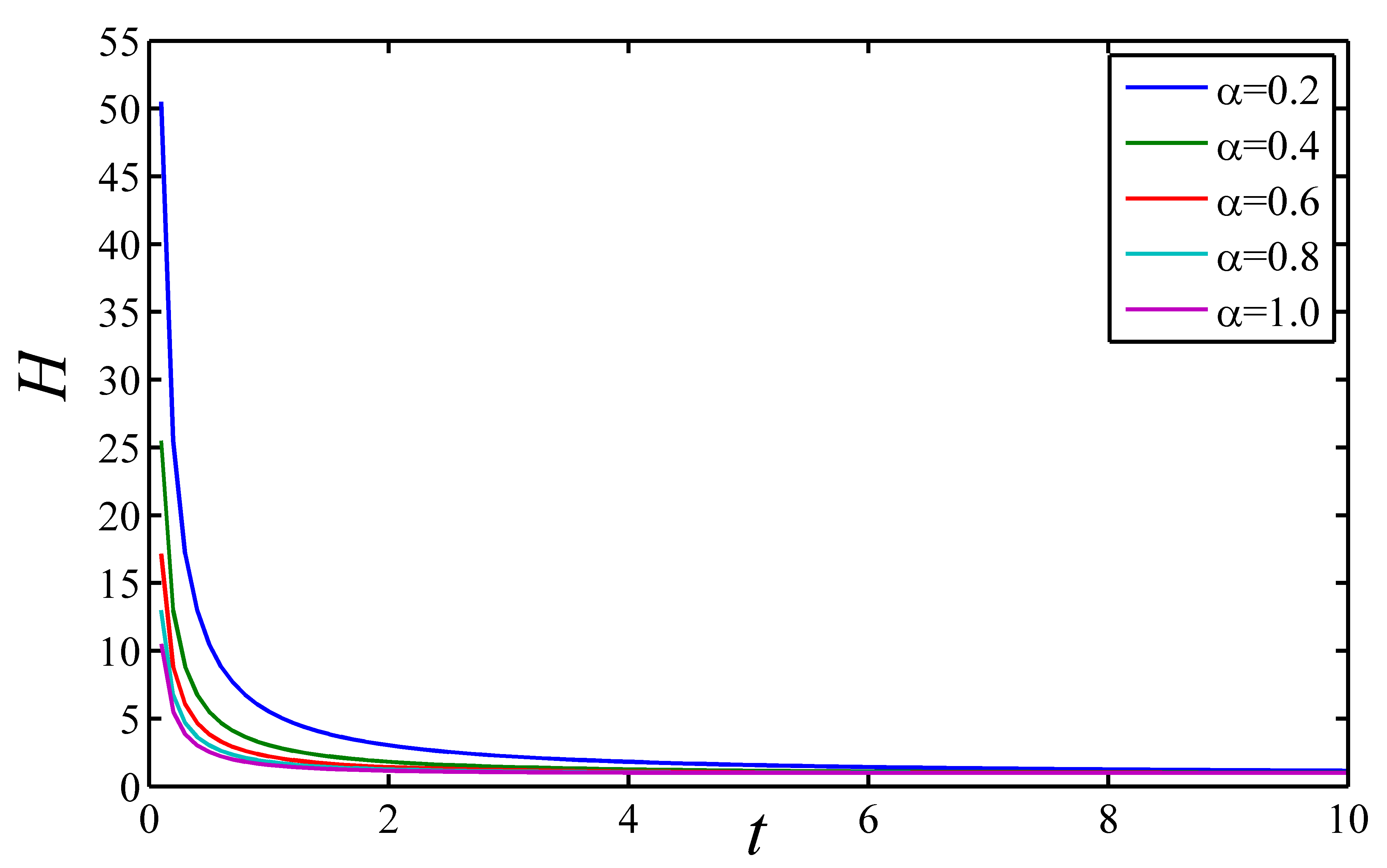}
  \caption{Variation of Hubble parameter against time for different $\beta$}\label{fig14}
\endminipage\hfill
\minipage{0.32\textwidth}%
  \includegraphics[width=60mm]{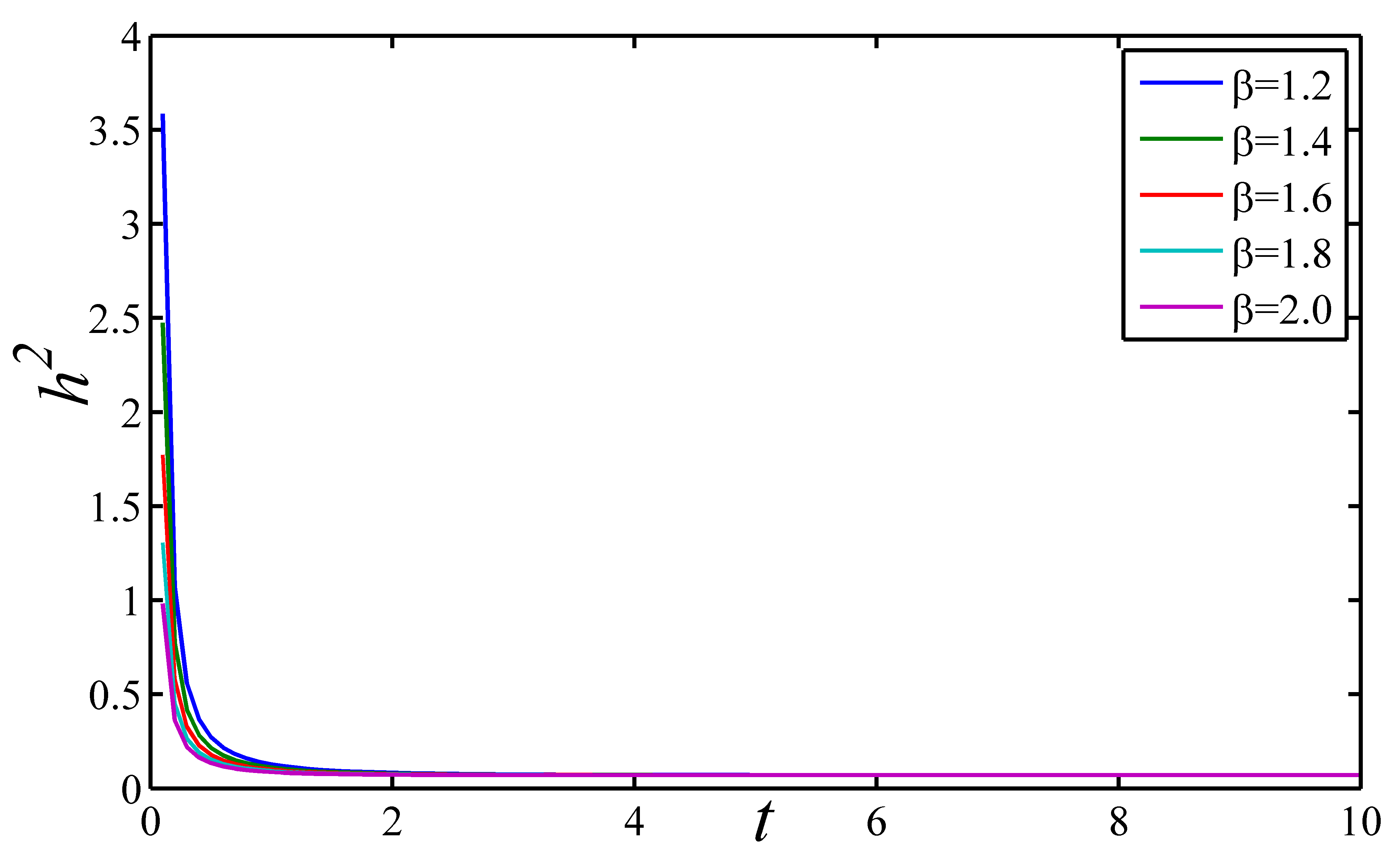}
  \caption{Variation of magnetic flux $h^2$ against time for $\lambda=0.1$, $m=0.5$, $B_c=60$ and different $\beta$}\label{fig15}
\endminipage
\end{figure}

\begin{figure}[ht!]
\minipage{0.32\textwidth}
  \includegraphics[width=60mm]{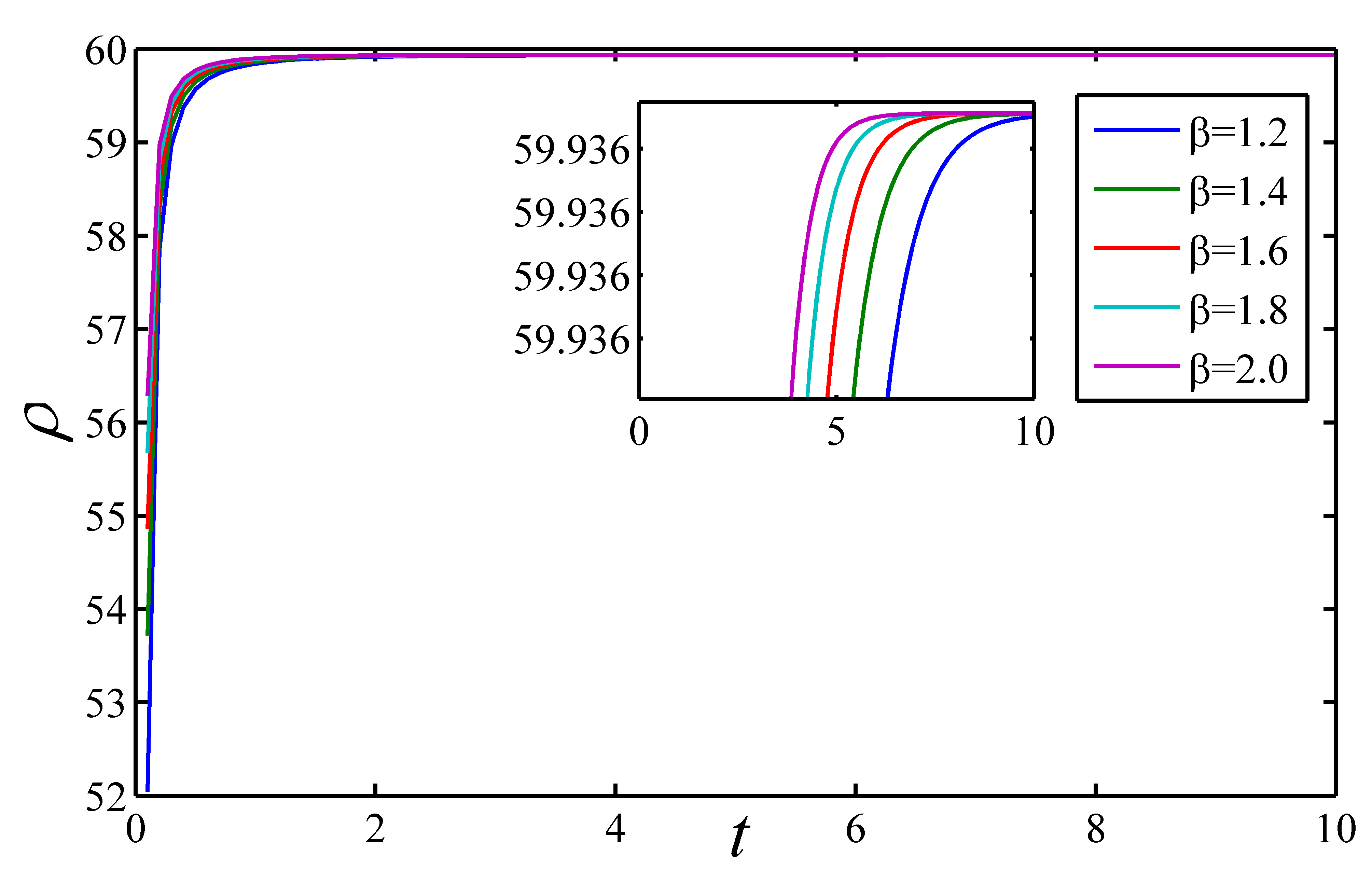}
\endminipage\hfill
\minipage{0.32\textwidth}
  \includegraphics[width=60mm]{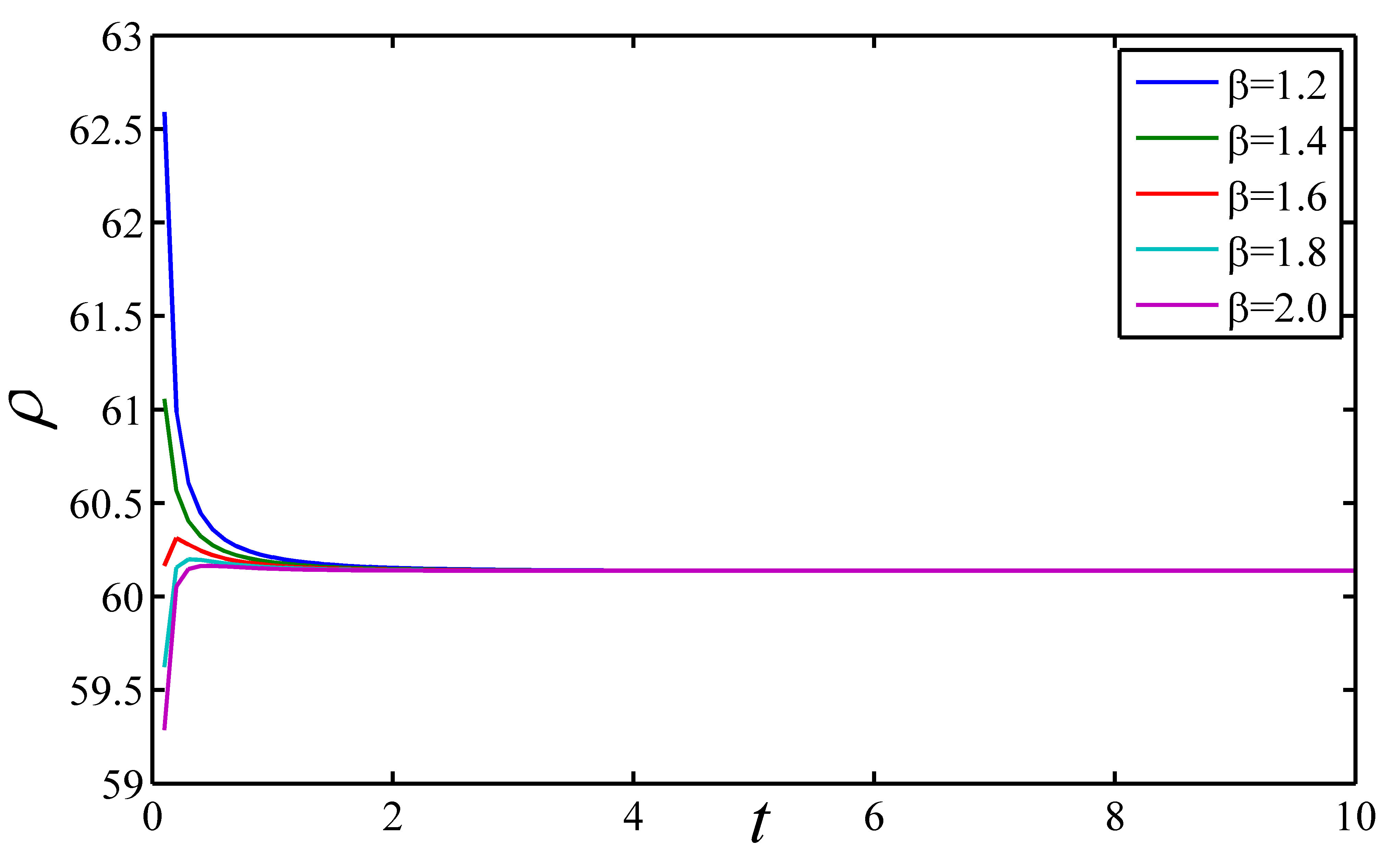}
\endminipage\hfill
\minipage{0.32\textwidth}%
  \includegraphics[width=60mm]{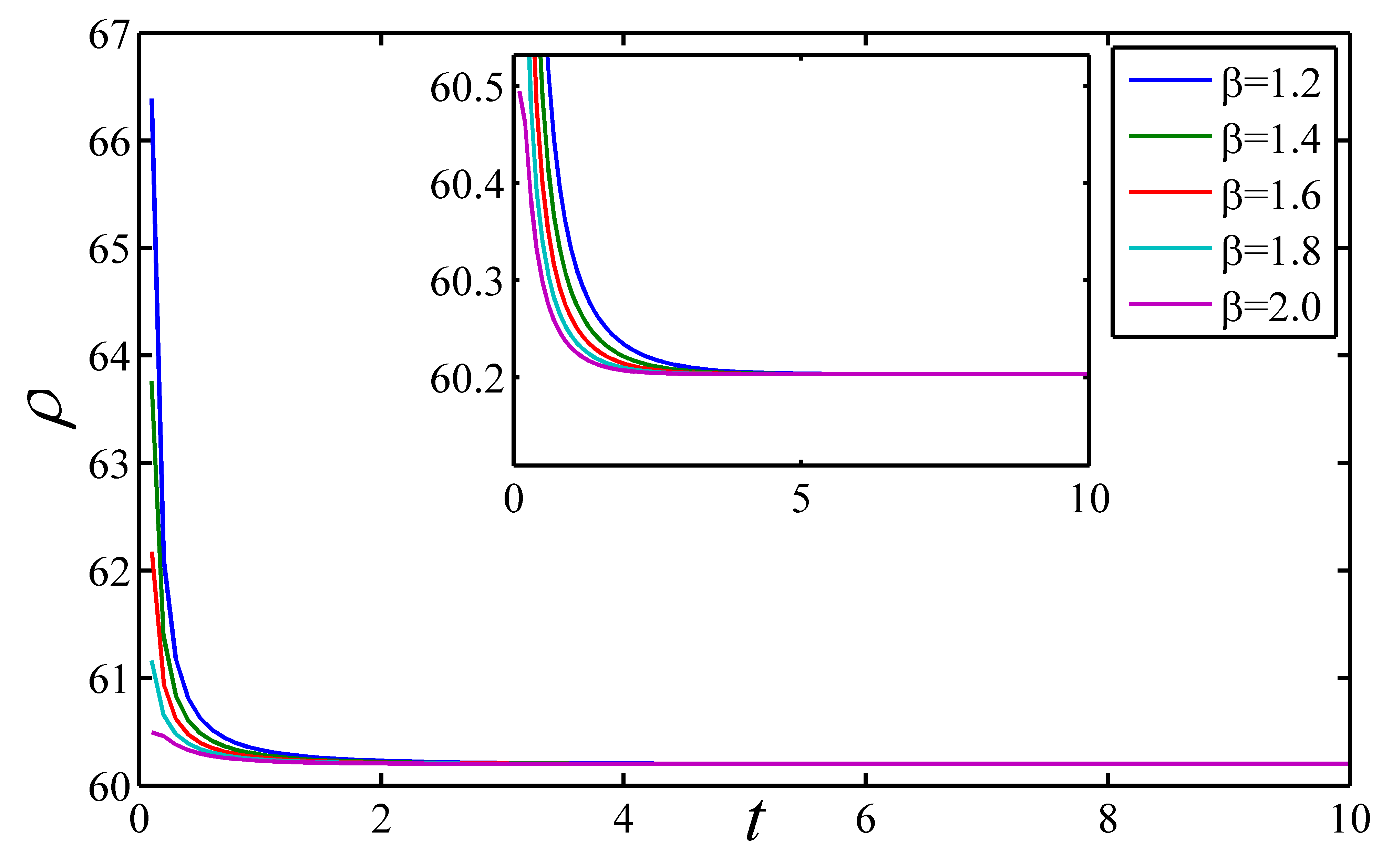}
\endminipage
\caption{Variation of energy density against time for $\lambda=0.1$, $B_c=60$ and different $\beta$ with $m=0.5$, $m=2.5$ and $m=3.6$ respectively}\label{fig16}
\end{figure}

\begin{figure}[ht!]
\minipage{0.32\textwidth}
  \includegraphics[width=60mm]{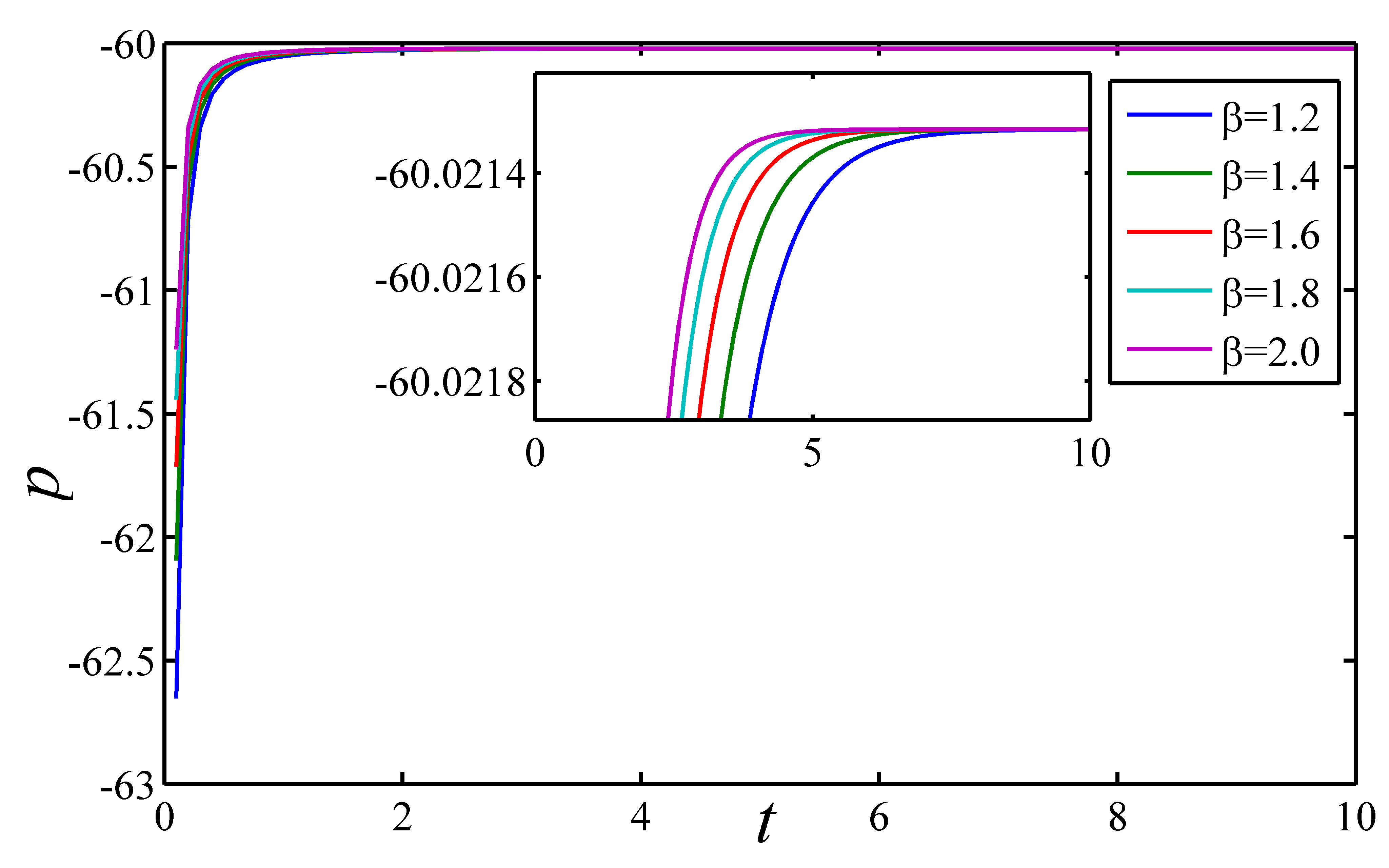}
\endminipage\hfill
\minipage{0.32\textwidth}
  \includegraphics[width=60mm]{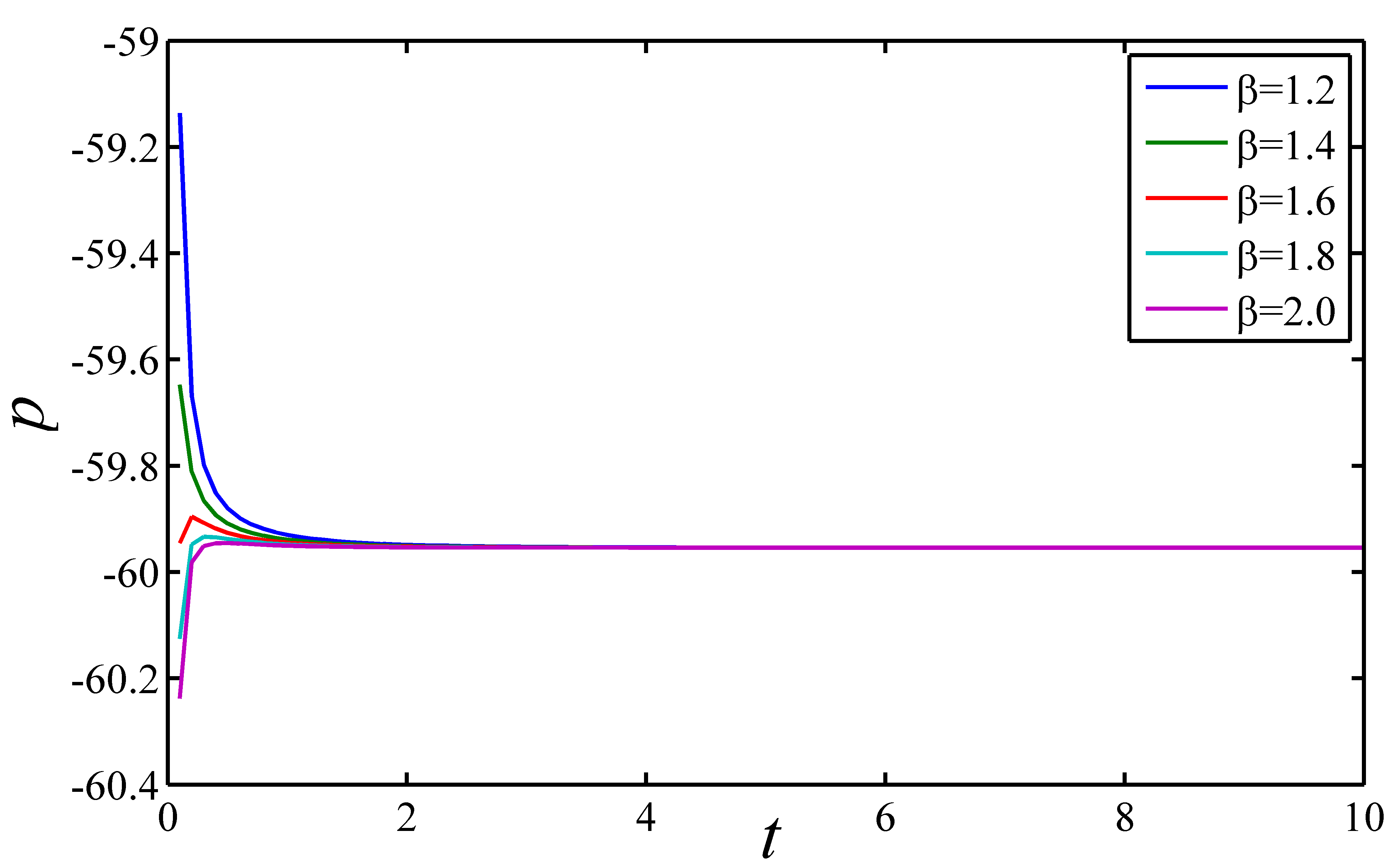}
\endminipage\hfill
\minipage{0.32\textwidth}%
  \includegraphics[width=60mm]{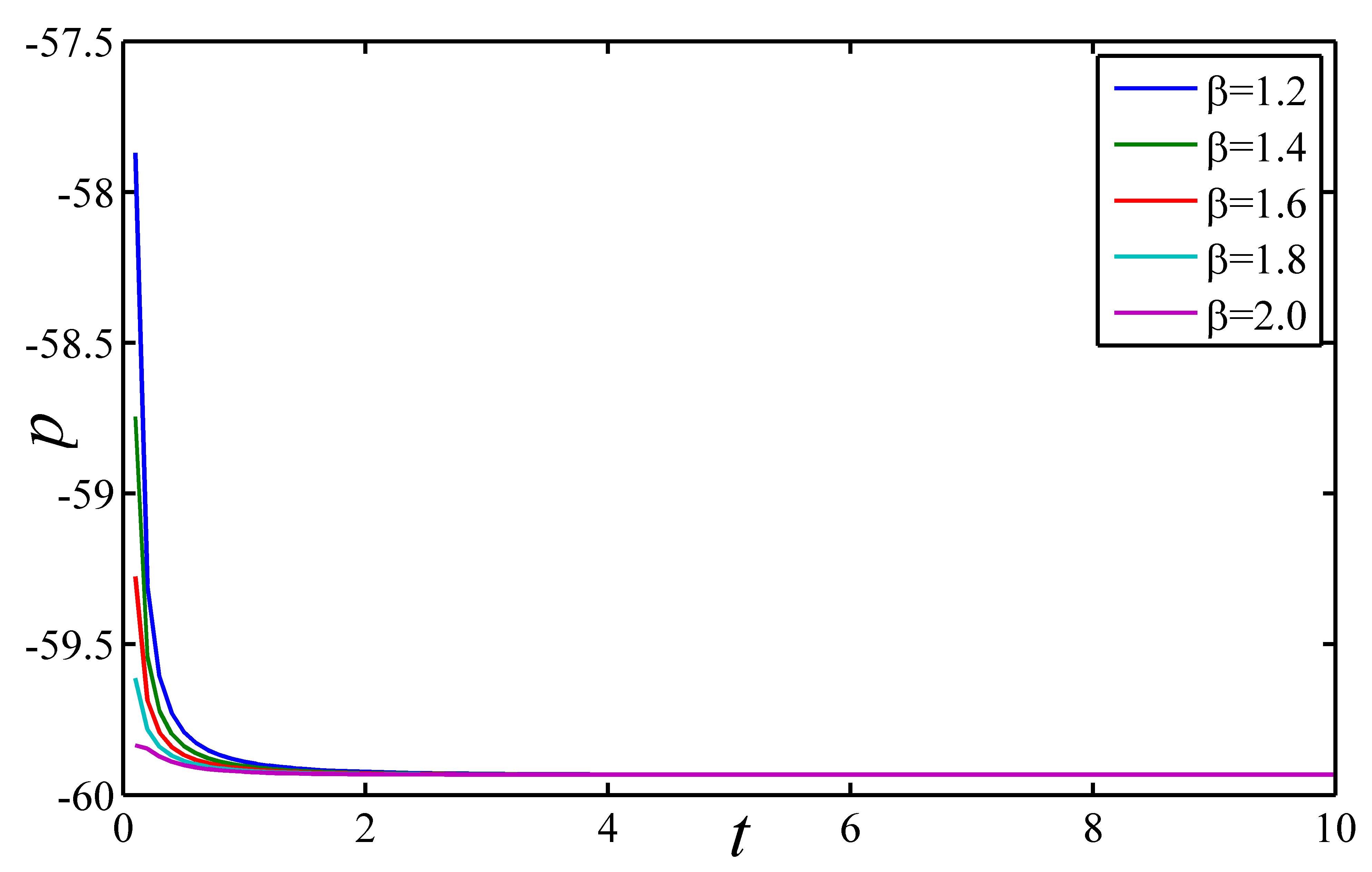}
\endminipage
\caption{Variation of pressure against time for $\lambda=0.1$, $B_c=60$ and different $\beta$ with $m=0.5$, $m=2.5$ and $m=3.6$ respectively}\label{fig17}
\end{figure}

\begin{figure}[ht!]
\minipage{0.32\textwidth}
  \includegraphics[width=60mm]{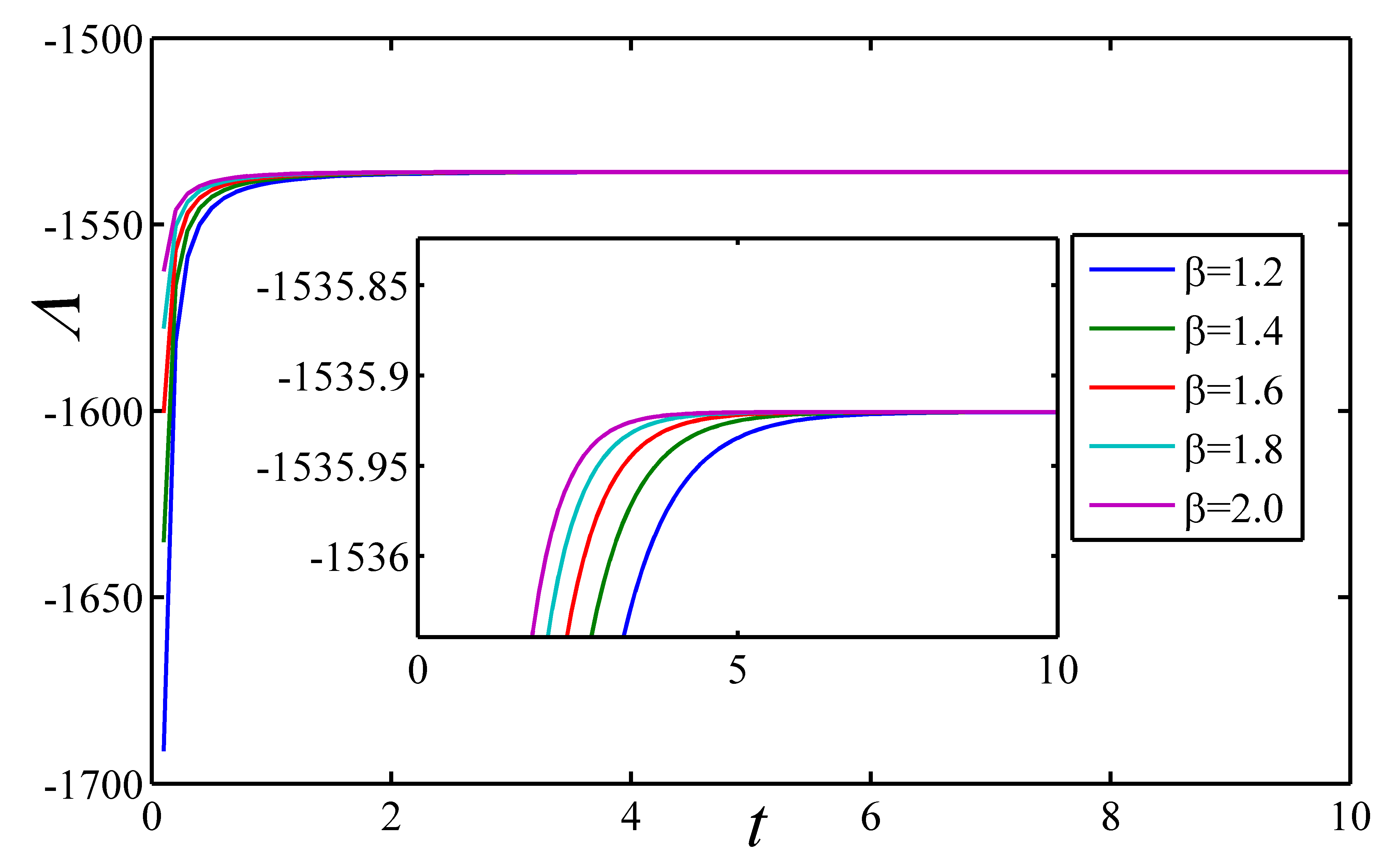}
\endminipage\hfill
\minipage{0.32\textwidth}
  \includegraphics[width=60mm]{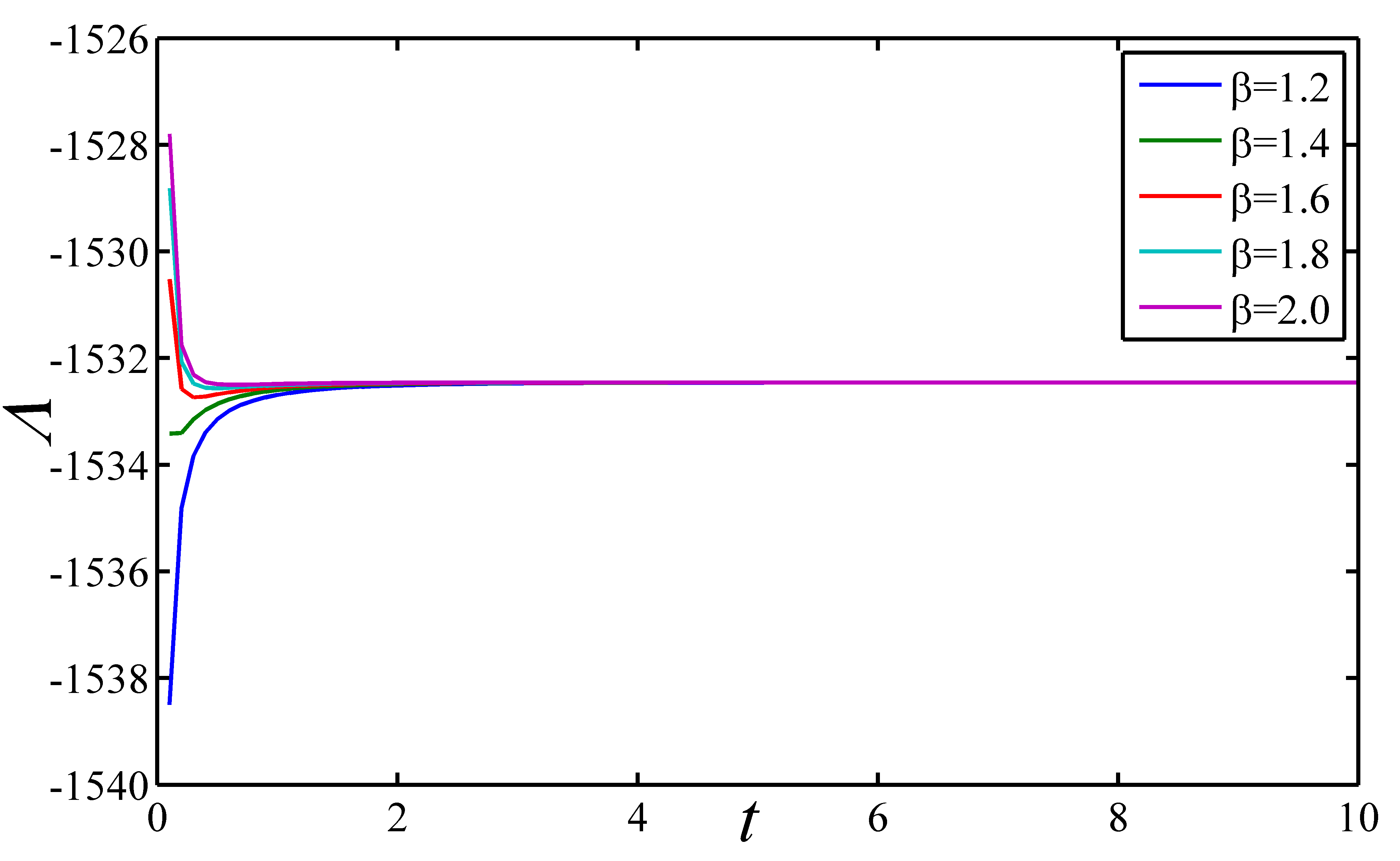}
\endminipage\hfill
\minipage{0.32\textwidth}%
  \includegraphics[width=60mm]{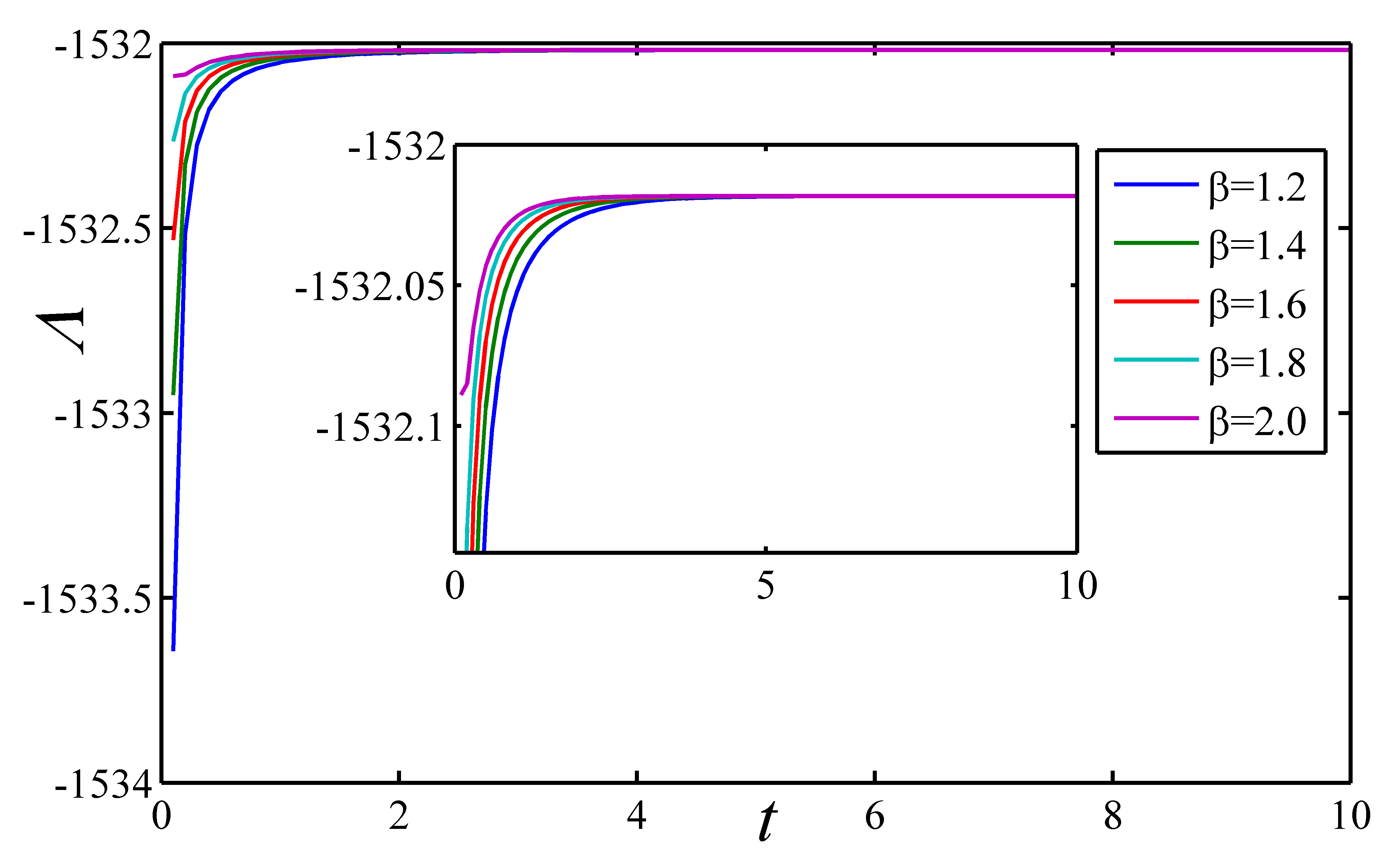}
\endminipage
\caption{Variation of cosmological constant against time for $\lambda=0.1$, $B_c=60$ and different $\beta$ with $m=0.5$, $m=10$ and $m=240$ respectively}\label{fig18}
\end{figure}

The variation of deceleration parameter, Hubble parameter and magnetic flux against time is presented in Figure \ref{fig13}, Figure \ref{fig14} and Figure \ref{fig15} respectively. Here we noticed that, $q\in(0,-1)$ for $\beta \in (0,1]$ and $q\in(-1,1)$ for $\beta \in (1,2)$. In the interval of $\beta \in (0,1]$, the deceleration parameter is negative valued whereas for $\beta \in (1,2)$ it takes values from positive to negative, which means that in the interval phase transition takes place. Hubble parameter is a positive, decreasing valued function of time and approaches to zero with the increment of time.  The magnetic flux $h^2$ is positive and decreasing function of time for $n \in (0,1)$ and provided $\beta$. As we are interested in the case of phase transition, all the physical parameters are presented graphically with $\beta \in [1,2]$. The variation of energy density against time is depicted in the Figure \ref{fig16}. From equation (\ref{50}), one can noticed that, $\rho\rightarrow B_c$. In this case, we would like to mention that, the approach of energy density towards $B_c$ is different for different interval of $n$ and $\beta \in [1,2]$. For $n\in(0,2]$ and $n\geq 3.6$, $\rho$ approaches to $B_c$ from left and right of $B_c$ respectively. In case of   $\beta \in [1,2]$ and $n\in [2,3.6]$, $\rho\rightarrow B_c$ from either side of $B_c$. As a representative case, we choose three different values of $n$ i.e.\ $n= 0.5, 2.5, 3.6$ and $\beta \in [1,2]$ (see Figure \ref{fig16}). Pressure profile also have similar qualitative behaviour as that of energy density but here $p\rightarrow -B_c$ (see Figure \ref{fig17}). The cosmological constant is negative valued function of time. We have observed that, for $n\in(0,2)$, $n\in[2,235]$ and $n>235$, $\lambda\rightarrow -(8\pi+4\mu)B_c$ from left, either side and right of $-(8\pi+4\mu)B_c$ respectively.  As a representative case, we choose three different values of $n$ i.e.\ $n= 0.5, 10, 240$ and $\beta \in [1,2]$ (see Figure \ref{fig18}). The other physical quantities like volume, shear scalar, expansion scalar have the similar qualitative behaviour as that of model-I.

\section{Conclusion}\label{sec-IV}

In this paper we have presented a cosmological model with the linear form (i.e. $f(R,T)=R+2f(T)$) of $f(R,T)$ gravity with magnetized strange quark matter. The exact solutions of the field equations are obtained using three different forms of DP.\\
The findings of this work are quite convincing, and the conclusions can be drawn as follows:
\begin{itemize}
\item The deceleration parameter shows a signature flipping for a universe which was decelerating in past and accelerating at present epoch. Therefore, the DP is a most physically justified parameter to analyse the solution of cosmological models. Moreover, we observed that the first model with a bilinear DP represents a transition of universe from early decelerating phase to a recent accelerating phase. In second case, the universe lies at an accelerating phase. The third model shows a transition of universe for $\beta> 1$ and again lies at an accelerating phase for $\beta\leq 1$ (See Fig. \ref{fig13}). Summing up the results, it can be concluded that, the deceleration parameter plays a vital role in account of accelerated expansion of the universe. The model with time varying deceleration parameter represents an expanding universe in accelerated phase.
\item  Each model represents an accelerated expansion of the universe as $q<0$ and $V\rightarrow \infty $ at $t\rightarrow \infty$. It can be noted that, in the early universe the magnetic flux has more effects and its effects gradually reduces in later stage.
\item The bag constant plays an important role in the expansion of the universe. One can see form our discussed modes that in each model $p \rightarrow - B_c$ for late time. Here the negative sign indicate the expansion of the universe in late time. Thus larger value of bag constants leads to larger expansion.
\item The pressure and energy density of each model approaches to bag constant in negative and positive way at $t\rightarrow \infty$ i.e. $p\rightarrow -B_c$ and $\rho\rightarrow B_c$ at $t\rightarrow \infty$. As per the observation, the negative pressure is due to the dark energy in the context of accelerated expansion of the universe. So the strange quark matter along with magnetic field gives an idea of existence of dark energy in the universe and supports the observations of the type Ia Supernovae \cite{riess98}. Also our results agree with the study of Akta\c{s} and Ayg\"{u}n \cite{Aktas17}. They researched MSQM distribution in $f(R,T)$ gravity and found $p=-\rho$ dark energy model for $t\rightarrow \infty$.
 \item  The scalar expansion shows that the expansion rate is faster at the beginning and becomes slow down in later stage.
\item In the discussed three models, one can observe that, the bag constant involves in the three physical parameters namely energy density, pressure and cosmological constant.  So $B_c$ does not affect the shear parameter and magnetic flux. In all the discussed models at late time the energy density, pressure and cosmological constant maintain a constant value equal to $B_c$. That means more negative pressure and more negative cosmological constant, which may responsible for accelerated expansion of the universe.  In late time the energy density is also maintain the same for all the discussed models here.
\item Since in each case, the shear scalar $\sigma^2 \neq 0$ and average anisotropy parameter gives a constant value i.e $\Delta=\frac{2(n-1)^2}{(n+2)^2} \neq 0$. Hence, the models obtained here with three different deceleration parameter  represents expanding, shearing and an anisotropic universe.
\end{itemize}

\begin{acknowledgements}
PKS, PS acknowledge DST, New Delhi, India for providing facilities through DST-FIST lab, Department of Mathematics,
where a part of this work was done.
\end{acknowledgements}

\end{document}